\documentclass[12pt]{article}
\usepackage{amsmath}
\usepackage{amssymb}
\usepackage{latexsym}
\usepackage{epsfig}
\usepackage{graphicx}
\usepackage{slashed}

\usepackage{natbib}

\usepackage{amssymb}
\usepackage{euscript}

\catcode`\@=11 \@addtoreset{equation}{section}

\def\0{\nonumber}

\def\Tr{{\rm Tr}}
\def\exp{{\rm exp}}

\def\Tr{ \hbox{\rm Tr}}

\def\dag{{}^{\dagger}}

\newcommand{\Aa}{{\cal A}}
\newcommand{\FF}{{\cal F}}

\newcommand{\CC}{{\cal C}}

\newcommand{\OO}{{\cal O}}
\newcommand{\HH}{{\cal H}}

\newcommand{\NN}{{\cal N}}
\newcommand{\TT}{{\cal T}}

\newcommand{\DD}{{\cal D}}

\newcommand{\VV}{{\cal V}}
\newcommand{\wt}{\widetilde}

\newcommand{\ben}{\begin{eqnarray}\displaystyle}
\newcommand{\een}{\end{eqnarray}}

\newcommand{\al}{\alpha}

\newcommand{\la}{\lambda}

\newcommand{\ve}{\varepsilon}
\newcommand{\De}{\Delta}
\newcommand{\pdx}{\frac{\partial}{\partial x}}

\newcommand{\dda}{{\dot{\alpha}}}

\newcommand\ee{\end{eqnarray}} 
\newcommand\be{\begin{eqnarray}}
\newcommand\ba{\begin{array}} 
\newcommand\ea{\end{array}}
\newcommand\eeq{\end{equation}} 
\newcommand\beq{\begin{equation}}
\newtheorem{HW}{Problem}[section]

\def\res#1{ { \hat {#1}}}
\def\bar#1{\overline #1}
\newcommand{\dsl}{\raise.15ex\hbox{/}\kern-.57em\partial}
\def\aslash{\raise.15ex\hbox{/}\mkern-14mu A}
\newcommand{\pslash}{\raise-.15ex\hbox{/}\mkern-9.5mu p}

\newcommand{\Dslash}{\hbox{/\kern-.6000em D}}
\newcommand{\dslash}{\,\raise.15ex\hbox{/}\mkern-13.5mu D}

\newcommand\mathC{\mkern1mu\raise2.2pt\hbox{$\scriptscriptstyle|$}
        {\mkern-7mu\rm C}}              

\newcommand\bi{\begin{itemize}}
\newcommand\ei{\end{itemize}}

\begin{document}


\vspace{.1in}
\begin{center}
{\large\bf On emergence in gauge theories at the 't Hooft limit} \\
\vspace{5mm}
\end{center}
\vspace{0.1in}

\begin{center}
Nazim Bouatta$^1$ and Jeremy Butterfield$^2$ \vspace{7mm}

$^1$ {\it Darwin College and DAMTP,
University of Cambridge, CB3 0WA, UK}\vspace{5mm}

$^2$ {\it  Trinity College, Cambridge,  CB2 1TQ, UK}

\end{center}


\vspace{2mm}

\begin{center} {\tt  N.Bouatta@damtp.cam.ac.uk,  jb56@cam.ac.uk} \end{center}
\vspace{0.1in}

\begin{abstract}
\noindent Quantum field theories are notoriously difficult to understand, physically  as 
well as philosophically. The aim of this paper is to contribute to a better conceptual 
understanding of gauge quantum field theories, such as quantum chromodynamics, by 
discussing  a famous physical limit, the 't Hooft limit,  in which the theory concerned 
often simplifies.

The idea of the limit is that the number $N$ of colours (or charges) goes to infinity. The 
simplifications that can happen in this limit, and that we will consider, are: (i) the 
theory's Feynman diagrams can be drawn on a plane without lines intersecting  (called 
`planarity'); and (ii) the theory, or a sector of it, becomes integrable, and indeed 
corresponds to a well-studied system, viz. a spin chain. Planarity is important because it 
shows how a quantum field theory can exhibit extended, in particular string-like, 
structures; in some cases, this gives a connection with string theory, and thus with 
gravity.

Previous philosophical literature about how one theory (or a sector, or regime, of a 
theory) might be emergent from, and-or reduced to, another one has tended to emphasize 
cases, such as occur in statistical mechanics, where the system before the limit has 
finitely many degrees of freedom. But here, our quantum field theories, including those on 
the way to the 't Hooft limit, will have infinitely many degrees of freedom.

Nevertheless, we will show how a recent schema by Butterfield and taxonomy by Norton  apply 
to the quantum field theories we consider; and we will classify three physical properties 
of our theories in these terms. These properties are planarity and integrability, as in (i) 
and (ii) above; and the behaviour of the beta-function reflecting, for example, asymptotic 
freedom.

Our discussion of these properties, especially the beta-function, will also relate to 
recent philosophical debate about the propriety of assessing quantum field theories, whose 
rigorous existence is not yet proven.

 \end{abstract}
\newpage

\tableofcontents

\section{Prospectus}\label{prosp}

 Physical theories often simplify, and so can be better understood, in some limit. A 
familiar example is the thermodynamic limit in statistical mechanics, which takes the 
number $N$ of constituents to infinity (while actually $N \approx 10^{23}$). Here, we hope 
to contribute to a better conceptual understanding of the heuristic formulation of gauge 
quantum field
theories, by discussing their 't Hooft limit:  in which the theory concerned often 
simplifies.

This simplification is very surprising because the limit is defined, for quantum 
chromodynamics (QCD) and other gauge theories, by taking the number $N$ of colours, which 
in QCD is actually three, to infinity: thus passing from $N = 3$ through apparently very 
complex theories, with say googol ($10^{100}$) colours.  Just as in thermodynamics  the 
limit $N \rightarrow \infty$ is ``controlled'' by requiring that a quantity like the 
density remain fixed (so that the volume also goes to infinity): so also here, the limit is 
controlled by requiring conditions on the product $\lambda := g^2N$ (of the square of the 
coupling constant $g$ with $N$). $\lambda$ is called `the 't Hooft coupling', and the 
condition is `being fixed'.

We will connect the physics of these limits to philosophical discussions of one theory (or 
a sector, or regime, of a theory) being emergent from, and-or reduced to, another one. 
Besides, our physical and philosophical claims are independent of much of the details of 
QCD. They apply to other non-abelian gauge fields, i.e. with a gauge group different from 
QCD's $SU(N)$, e.g. $SO(N)$ or $U(N)$.

The 't Hooft limit is a rich subject, with many aspects which we cannot pursue here. One 
main one, which we will study in a companion paper, is that this limit sheds light on the 
connection between quantum field theories and gravity---a connection much studied under the 
label `AdS/CFT'. Here we will instead be concerned with three aspects, occurring in two 
theories, viz. QCD and a supersymmetric cousin of it. These aspects are:\\
\indent (1): In both theories, we will be concerned with the high-energy behaviour of 
interaction strengths as described by the beta-function, i.e. asymptotic freedom or 
conformal invariance (Sections \ref{free} and \ref{beta}). As we will discuss in Section 
\ref{rigour}, this aspect is central to the question whether the theory rigorously exists. 
\\
\indent (2): We will discuss emergent planarity and related string-like structures, first 
for  QCD (Section \ref{planar}) and then for the supersymmetric cousin (Section 
\ref{anom}). This aspect relates closely to the string-gravity connection, AdS/CFT; and to 
the appearance of a classical regime.\\
\indent (3): We will discuss emergent integrability, in the supersymmetric cousin of QCD 
(Sections \ref{symmies} to \ref{chains}). This aspect relates closely to whether the theory 
is tractable for calculation and-or mathematical analysis.

Since in the `t Hooft limit, the parameter $N$ is the number of colours, the situation is 
different from those considered in philosophical discussions of statistical mechanics. 
There, the parameter is often the number of degrees of freedom, or a similar notion like 
the number of constituent particles; while here, in field theories, the number of degrees 
of freedom is infinite.

But despite this difference, our three aspects (1)-(3) will fit in with those discussions. 
In particular, we will classify the aspects using a recent schema of Butterfield's and a 
taxonomy of Norton's, about the degree of ``meshing'' between what holds good at the limit 
$N = \infty$, and what holds good before the limit, at finite $N$. This meshing makes the 
limit valuable as a framework for studying the finite $N$ theories: which are those of 
physical interest, since no one believes that the number of colours is in fact infinity. 
Thus one even envisages making an expansion in $1/N$ about the limiting theory, and 
then---again: very surprisingly---getting a physical result by substituting in the actual 
(finite!) value of $N$, even a very small one like 3. 

We now turn to a Section-by-Section prospectus.\\


Section \ref{new2} sets the scene and includes a summary (Section \ref{classify}) of how 
our three aspects (1)-(3) will be classified in our two theories. So the reader in a hurry 
can even stop after Section \ref{new2}. In more detail: in Section \ref{landscape}, we 
will: (i) present Butterfield's schema for inter-theoretic relations, which allows 
emergence, i.e. novelty of theoretical structures, to be combined with reduction, by taking 
a limit; (ii) present Norton's taxonomy---in fact, a trichotomy---of degrees of meshing at 
the limit; and then (iii) compare the schema and trichotomy. In Section \ref{trichUS}, we 
will discuss: (i) whether our theories, and others like them, rigorously exist; (ii) how 
this issue is affected by two remarkable properties of the theories, viz. asymptotic 
freedom and conformal invariance; and  (iii) how this issue affects the philosophical 
discussion. Then, using these discussions' conclusions, Section \ref{classify} sums up how 
our three aspects (1)-(3) will be classified.

Then in Section \ref{tHooft}, we  sketch  some relevant aspects of gauge theories, 
especially QCD.  The main point will be the surprising simplification. QCD is described by 
a gauge group $SU(3)$, where 3 is the number of colours: the gauge fields are described by  
3 $\times$ 3 matrices. So one naturally expects that theories using a  value of $N$ higher 
than 3, i.e. using larger matrices, will be increasingly complex. But in 1974, `t Hooft 
discovered that the theory simplifies at $N = \infty$. In the perturbation 
(Feynman-diagram) expansion, only those diagrams that can be drawn on a two-dimensional 
plane without intersecting lines (called `planar diagrams') remain. This is associated with 
the appearance of string-like structures; and with the perturbation expansion being in a 
certain sense a complete representation of the theory.\footnote{As we shall see, the main 
rationale for focussing on QCD is that---apart from the historical point that `t Hooft 
introduced his limit for QCD---it is asymptotically free; and this is related to the 
theory's rigorously existing  (cf. respectively  Section \ref{free} and Section 
\ref{trichUS}).}

In Section \ref{limitqft}, we describe the novel features and mathematical structures that 
can occur in the $N = \infty$ limit. But we `change horses', i.e. we consider a different 
theory. For in this limit, QCD remains  a complicated theory. So we emphasize, as the 
physics literature does, a simpler theory, again in this limit: called `maximally 
supersymmetric Yang-Mills theory'. Despite the long name, this is simpler than QCD! That 
is: the details of its dynamics are easier to study. We will dub it `SYM'.\footnote{As we 
shall see in Sections \ref{SYM} and  \ref{beta}, this simplicity arises from a high degree 
of symmetry, especially conformal invariance as given by a vanishing beta-function. 
Besides, field theories with a supersymmetry which is less than maximal (i.e. less than in 
SYM) have been shown to be in some respects easier to study than non-supersymmetric 
theories, and in particular to exhibit analytically important  features of quantum
gauge theories, such as confinement and a mass gap: e.g.
Seiberg and Witten (1994).} SYM is also guessed and hoped to be one of the essential models 
for understanding more complicated gauge theories, including QCD itself. So much so that 
nowadays SYM is sometimes called `the harmonic oscillator of
quantum field theory'.

We emphasize that our discussion of SYM does not require that nature actually be 
supersymmetric; just as our discussing the `t Hooft limit does not require that nature 
actually have infinitely many colours. More positively, we believe that foundational work 
can, and even should sometimes, consider theories because they shed light on conceptual 
questions and-or are analytically tractable: and we submit that both these reasons justify 
considering the 't Hooft limit of QCD and SYM. Indeed, these reasons also justify studying 
supersymmetry more generally. A good example of this is the beta-function itself: although 
in QCD we can only calculate this perturbatively, there are some supersymmetric theories 
whose beta-function has been computed exactly (Novikov et al. (1983), Shifman (2012, pp. 
531-533)).  (Of course, considering such  theories echoes the widespread acceptance of the 
Wilsonian perspective on  renormalization, with its use of a space of theories, connected 
by the renormalization group flow.)

In Section \ref{inty}, we arrive at integrability in the 't Hooft limit. That is: we 
discuss how, thanks to planarity, certain physical aspects of SYM are mapped into 
integrable spin chains. Here we connect with an old dream of quantum field theory: to 
calculate analytically the mass-spectrum of a theory, i.e. the masses of its particles such 
as a proton, as a function of the parameters of the theory, such as coupling constants and 
the energy-scale. SYM is conformally invariant: that is, roughly speaking, invariant under 
a transformation that changes scales but preserves angles. This makes the theory have no 
massive particles. But there is an analogue of the mass that one can aspire to compute: 
namely, the scaling dimension of local operators. (It is also an even closer analogue of 
the critical exponents occurring in statistical mechanics' description of phase 
transitions.) The idea is that the correlation function of such an operator, i.e. the 
correlation (the average of the product) of its values at two different spacetime points 
(as given by an expectation value of a product), falls off with some power $\Delta$ of the 
spatiotemporal distance between the two points. This $\Delta$ is the scaling dimension: it 
includes a quantum addition (called `anomalous dimension') to the term $\Delta_0$ obtained 
by classical dimensional analysis. And it has recently been shown that $\Delta$ can be 
calculated by analysing the associated spin chain systems: a special case of the old 
dream.\footnote{This hope is called the `old dream' at the start of the excellent recent 
review of integrability in this context, by Beisert et al. (2012), to which (together with 
Minahan (2012)) our Sections \ref{limitqft} and \ref{inty} are indebted: cf. also Polyakov 
(1987, Chapters 8-10). Although we have postponed to our companion paper discussion of 
string theory as a theory of gravity, we should stress that these recent advances arose 
from examining the connection between quantum field theories and string theory, more 
specifically AdS/CFT.}

Finally in Section \ref{nonplanar}, we briefly discuss the integrability of SYM {\em 
before} the $N = \infty$ limit. Considering the situation at finite $N$ is important, 
because someone might object to the results of Section \ref{inty}: `All very well, but any 
correct quantum field theory of the actual world surely has finite $N$.' This echoes a 
familiar objection often urged by philosophers of statistical mechanics about the appeal to 
the thermodynamic limit, in which the number of constituent particles is taken to be 
infinity, in order to explain phase transitions. In effect, they object: `my kettle has 
finitely many atoms in it!'. There are of course parallel misgivings about other 
explanations using infinitary models of finite systems.

We believe the main answer to this objection is that even before the limit, there is a 
version of the phenomenon at issue: in the familiar discussions,  phase transitions---and 
here, asymptotic freedom, conformal invariance, planarity and string-like structures. That 
version may be ``weaker'' or ``approximate'', in various senses, by comparison with the 
phenomenon at the limit. But it is nevertheless real. We also take this answer to be 
uncontroversial: we think it is part of how physicists construe such infinite 
idealizations---albeit usually implicitly, rather than explicitly. We discuss this further 
in Sections \ref{landscape} and \ref{trichUS}.

But we should admit here that the situation for integrability is a bit different. As we 
shall see in Section \ref{nonplanar}, not much is at present known about integrability at 
finite $N$, i.e. on the way to the `t Hooft limit. So for physics, this is a 
straightforward case of work for the future. For philosophy, it illustrates the point that 
philosophical morals drawn from physics need to beware of the limitations of the present 
state of knowledge. Finally,  Section \ref{concl} mentions two further topics to be pursued 
elsewhere.

\section{At the limit vs. before the limit, in gauge theories}\label{new2}

\subsection{A schema and a trichotomy}\label{landscape}
Before embarking on the details, we should connect what follows with recent philosophical 
discussion of inter-theoretic relations, especially the contrast between {\em at} the limit 
and {\em before}, or {\em on the way to}, the limit. Perceptive recent work includes 
Callender and Menon (2012), who discuss phase transitions; and Norton (2012) who discusses 
examples from both thermal physics and geometry. We will connect just with the trichotomy
proposed in Norton's paper: which, we claim, will develop a schema from our own previous 
discussions (Butterfield (2011,
especially Section 3.1-3.2, pp. 1073-1076), Bouatta and Butterfield (2011)).

\paragraph{2.1.A The schema} We proposed a schema and a mnemonic notation, for 
inter-theoretic relations. We wrote $T_b$ for the `better, bottom or basic' theory, and 
$T_t$ for the `tainted or top' theory. The schema was that (i) in some cases $T_t$ is 
deduced from $T_b$ (taken together with suitable auxiliary definitions), in some limit of a 
parameter; and (ii) although deduced, $T_t$ exhibits novel yet robust properties compared 
with those in $T_b$. Thus phase transitions  illustrate the schema: with $T_b$ taken as the 
statistical mechanics of $N$ constituents; $T_t$ as thermodynamics, taken as describing 
phase transitions in terms of singularities of thermodynamic quantities; and with the limit 
being the thermodynamic limit, $N \to \infty$. Butterfield (2011, Section 1, pp. 1066-1071) 
argued that this schema's combining (i) and (ii) amounted to reconciling reduction (taken 
as deduction {\em \`{a} la} (i)) with emergence (taken as novelty {\em \`{a} la} (ii)).

This paper will illustrate the same schema, in the following way.  $T_t$ will be either QCD 
or SYM (or even another gauge theory) in the 't Hooft limit. The two main novel yet robust 
properties (of course related to each other) are:  string-like structures as shown by 
planarity of Feynman diagrams (cf. Section \ref{planar}, and Section \ref{plandominate}), 
and integrability using spin chains (Section \ref{chains}). $T_b$ is of course QCD or SYM  
at finite $N$ (say, $N = 3$!). Thus again, we will see reduction and emergence 
reconciled.\footnote{Our companion paper on AdS/CFT discusses the corresponding theme, that 
gravity can be both reduced to and emergent from a gauge quantum field theory.}

\paragraph{2.1.B The trichotomy} Norton's recent analysis (2012) develops this schema. He 
is concerned to contrast three cases concerning the limit of a theory as a parameter $N$ 
goes
to infinity. We will first report his trichotomy of cases (2012, Sections 3.1-3.3), and 
compare it with our schema.  Then in Section \ref{trichUS}, we summarize how the three 
aspects of QCD and SYM that concern us are classified by his trichotomy, and how this sheds 
light on conceptual aspects of these gauge theories. In later Sections, this summary will 
of course get fleshed out.

The framework of discussion is that a physical theory describes systems, and their 
properties, in terms of physical quantities defined on the systems, and states which assign 
these quantities values (in quantum theories, expectation values). As the parameter $N$ 
varies from one system to another, there can be more or less ``meshing'' of the values of 
corresponding quantities on the systems (for an appropriate choice of states). And one can 
ask whether there is a mathematically well-defined system at the limit $N = \infty$. In 
Norton's examples, $N$ is the number of components, in some natural sense of `component'; 
e.g. the number of squares in a geometric lattice, or the number of atoms (or constituent 
particles) in a (model of) a sample of gas. So he illustrates  $N \rightarrow \infty$ by 
examples such as ever-finer lattices in geometry, or thermal physics' thermodynamic limit. 
Indeed, these examples are topical: philosophical discussion of limiting relations between 
theories often takes the parameter $N$ to be the number of components, in some natural 
sense.

Given this framework, we can now state Norton's three cases. In his first case, systems and 
properties ``mesh'' in the limit, as follows:\\
\indent (1): {\em Limit property and limit system agree}: There is a well-defined 
infinite-$N$ system, $\sigma(\infty)$ say.  It is usually obtained as a limit of the state 
spaces of finite, i.e. finite-$N$, systems $\sigma(N)$. And there is a property of the 
finite systems, usually given as the value of a quantity, that tends to a limit. Writing 
$f(N)$ for the quantity on the finite system (and so as short for $f(\sigma(N))$), one 
might write the value as  $v(f(N))$. (Here an appropriate  sequence of states $s_N$ is of 
course understood: $v(f(N))$ is short for $v(f(N), s_N)$). And this limit is the value of a 
well-defined quantity, $f(\sigma(\infty))$ say, on $\sigma(\infty)$ (where again, a 
limiting state $s_{\infty}$ is understood); and this quantity is itself a limit, in some 
natural sense, of the corresponding quantities on the finite systems. So in our notation: 
$\lim_{N \to \infty} v(f(N)) = v(f(\sigma(\infty)))$.

In Norton's second case, the meshing breaks down in that there is no infinite system; while 
in the third case, there is an infinite system, but the property considered for it does not 
match the limit of the property as defined on the finite systems. In more detail, we 
have:\\
\indent (2): {\em There is no limit system}: There is no infinite system; though there is, 
or may be, a limit of the property of the finite systems. In the notation just introduced: 
there is no $\sigma(\infty)$, but there is, or may be, a limit, $\lim v(f(N))$ i.e.  $\lim 
v(f(N), s_N)$.  (So again, the property is usually given as the value of a quantity for an 
appropriate sequence of states).\\
\indent (3): {\em Limit property and limit system disagree}: There is an infinite system 
$\sigma(\infty)$; (again, usually obtained as a limit of the state spaces of finite 
systems). But although the property of the finite systems (again: usually the value of a 
quantity) tend to a limit, this limit is {\em not} the value of the corresponding quantity 
on $\sigma(\infty)$. In our notation: $\lim v(f(N)) \neq v(f(\sigma(\infty)))$.

Note that each case is defined in terms of a single property  or quantity  $f$; or more 
exactly, in terms of a single family $f(N)$, with maybe also $f(\sigma(\infty))$ (with a 
sequence of states understood). But of course,  a typical theory considers many quantities, 
and so can illustrate more than one case. This happens in Norton's examples; and, as we 
will see in Section 2.1.C, it happens in ours.

Norton also proposes  that his trichotomy prompts a useful regimentation of the words 
`idealization' and `approximation', whose usage varies widely (his Section 2, pp. 208-211). 
Thus in case (1), Norton says that the infinite system is an {\em idealization}. The point 
of  an idealization is that its property (i.e. $v(f(\sigma(\infty)))$, the value of the 
quantity on the infinite system) gives an inexact description of what Norton calls the 
`target system'---i.e. the systems $\sigma(N)$ for realistic i.e. finite $N$. (Of course, 
$N$ might be very large, as in thermodynamics---cf. the kettle at the end of Section 
\ref{prosp}.)  Furthermore, the numerical agreement in the limit means that, although the 
description is inexact, it is  accurate {\em enough}. (The inexactness is just the 
difference between the limit of a sequence $\{ v_N \}_N$ of values, and a member $v_N$ of 
the sequence with a suitably large $N$ to be close enough to the limit.) And typically, its 
small inaccuracies are justified by it being much more tractable for calculation and 
manipulation than the finite-$N$ descriptions; (cf. Butterfield 2011, Section 3.3, p. 
1076-1082).

On the other hand,  Norton says that in case (2) there is a ({\em good}) {\em 
approximation}. That is: the limit $\lim v(f(N))$ gives a description of the target system 
that does not come from an idealization, and is inexact---but accurate {\em enough}, at 
least for high enough $N$. Finally,  Norton says that in case (3), we have a {\em poor 
approximation}, because the  property of the infinite system is not accurate enough a 
description of the target system.\footnote{Norton goes on to classify various limits, 
especially in thermal physics, e.g. the thermodynamic, continuum and Boltzmann-Grad limits, 
in terms of his cases. His main point is to argue that renormalization group methods are 
approximations, not idealizations; (his Sections 4.3, 5.2-3, pp. 219-223, 225-227). But we 
will not here try to systematically compare the 't Hooft limit with limits in statistical 
mechanics.}

\paragraph{2.1.C Comparison}  As we have seen, Norton is not concerned to define 
`emergence', or indeed reduction: unlike Butterfield. But it is natural to suggest that 
Norton's case (1) (`idealization') corresponds to Butterfield's reconciliation of emergence 
with reduction.  That is: the limit system's limit property is the novel, so emergent, 
property; while its being the limit of the properties of the finite-$N$ systems amounts to 
its being reduced.

In reply to this suggestion, we agree that a Nortonian case (1) might well be an example of 
emergence with reduction. But to say that his case (1) `corresponds' to emergence with 
reduction is too strong; for two reasons.

(1): There could be novel and emergent properties without their being properties of a limit 
system. There might be {\em no} limit system, as in Norton's case (2); so that the 
properties, despite their novelty and even importance, are uninstantiated. Besides, such a 
case could count as a reduction, owing to the novel property being a limit of the 
finite-$N$ properties. Thus we see that Butterfield's notion of emergence with reduction is 
in general logically weaker than Norton's case (1). No worries, say we: it reflects the 
flexibility of the idea of emergence.

(2): We should of course allow reduction to use properties other than the finite-$N$ 
version of the limit property. Recall our remark just after Norton's case (3): viz. that a 
theory considers many properties (quantities). Thus there might well be examples in which a 
limit property is reduced using finite-$N$ properties, i.e. quantities $q_1(N), q_2(N) 
...$, that are quite different from the finite cousin $f(N)$ of the limit property. (In 
this way, Butterfield's notion of reduction as deduction is more flexible than Norton's 
trichotomy, with its focus on a single sequence of properties: though of course, there is 
no conflict between our respective doctrines.) This second reason prompts another general 
comparison: as follows.

(3): It is natural to suggest that Norton's case (3) (`poor approximation') corresponds to 
a failure of reduction, at least in Butterfield's proposed sense, viz. as deduction. But 
again to say `corresponds'  is too strong, because of what we just said in (2). That is: 
there could be a deduction of the limit property, using quantities $q_1(N), q_2(N) ...$, 
that are quite different from its finite-$N$ cousins $f(N)$. Besides, this limit property 
could be important and even novel, and so emergent; even though it does not ``mesh'' with 
its finite-$N$ cousins $f(N)$.

So much by way of general comparison. In the rest of this paper, we will see two main 
examples of the scenarios of emergence with reduction, just envisaged in (1) to (3) above.

The first example corresponds to the first reason (1) above. It concerns QCD and SYM 
equally; and besides, several other quantum field theories. Even for a finite number $N$ of 
colours, the question whether an interacting quantum field theory rigorously exists is a 
subtle issue: which we will address in Section \ref{rigour}. But (1) above concerns, 
rather, whether the limit system---i.e. for us, a gauge theory with an infinite number of 
colours---rigorously exists. Happily, we will report in Section \ref{rigour} indications 
that for QCD and SYM, the limit theory does exist. But even if they do {\em not} exist, 
some of our three properties ((1) to (3) of Section \ref{prosp}) will exhibit emergence 
with reduction in our sense---as envisaged in (1) above. We will also suggest that this 
verdict accords with how theoretical physicists talk and write about the 't Hooft limit 
(e.g. Polyakov (1987, Chapter 8.1), Gross 1999, p. 588).\footnote{Later, we will briefly 
consider the suggestion that the limit theory is indispensable for understanding the 
finite-$N$ case. We note here that even if this is true, it does not mean that infinitely 
many colours of the
limit theory are physically real. Rather: if it is true, then we get a good understanding 
of a world, which is well but inexactly described by a heuristic theory with three colours 
i.e. QCD, by exploiting a cousin limiting theory.}

The second example corresponds to the second and third points, (2) and (3) above. We will 
see that SYM has been shown to have our third property, integrability, at the limit $N = 
\infty$, with a vivid representation using spin chains, by arguments that use---{\em not} 
integrability at finite $N$---but a different property: viz. the finite versions of our 
second property, planarity. That is: the arguments use the increasing dominance of Feynman 
diagrams that can be drawn on a plane. But despite using a different property, 
integrability is deduced, yet novel: emergence with reduction, again.

So far, this comparison has highlighted differences between Norton's discussion and our 
previous one. We should also note a similarity. Both Norton and Butterfield, like most 
previous philosophical discussions,
assume that the finite systems $\sigma(N)$ exist; and rightly so, for the examples they 
consider. But again: this involves two differences from our present study of the 't Hooft 
limit.\\
\indent (a): First: our parameter $N$, running through the positive integers, is not the 
number of degrees of freedom, since that is infinite in a field theory. Nor is it a kindred 
notion like the number of components, as in Norton's examples of the number of squares in a 
lattice, or the number of atoms in a sample. As we said in Section \ref{prosp}, our $N$ 
will be the number of colours: the $N$th theory describes its gauge field by an $N \times 
N$ matrix field, and so has gauge group $SU(N)$, rather than QCD's $SU(3)$.\footnote{Since 
for each $N$, the $N \times N$ matrix has $N^2 - 1$ independent components (since the group  
is $SU(N)$), there are $N^2 - 1$ gauge fields.}\\
\indent (b):  Second: since $N$ labels interacting quantum field theories, the question 
whether even the finite  $N$ theory rigorously exists will need to be addressed: cf. the 
next Section.

\subsection{The schema and trichotomy in QCD and SYM}\label{trichUS}
We turn to summarizing how what follows relates to Norton's trichotomy and Butterfield's 
reconciliation of reduction and emergence. In short: we will classify in terms of Norton's 
trichotomy, Section \ref{prosp}'s three properties that a quantum field theory---in an 
appropriate sequence of such theories, labelled by $N$---can have. The properties are, in 
short: (1) the high-energy behaviour of interaction strengths; (2) planar Feynman diagrams, 
and ensuing string-like structures; and (3) being integrable. We will see that in all 
cases, there is  good limiting behaviour: and in most cases, Norton's case (1), 
idealization.

We will need a preliminary discussion about rigour (Section  \ref{rigour}). Then in Section 
\ref{classify}, the classification of the  properties will be straightforward.

\subsubsection{Do gauge theories exist? Should we dive in?}\label{rigour}
We admit at the outset that, so far as is known, the theories we will consider, QCD and 
maximally supersymmetric Yang-Mills (SYM), are not defined, even at finite $N$, as 
precisely as a mathematician would require. For even at finite $N$, these are interacting 
quantum field theories in 3+1 dimensions, whose rigorous existence remains to be proven. So 
as we mentioned, there is a contrast with philosophical discussions of statistical 
mechanics. There, finite $N$  means finitely many degrees of freedom, and the theory or 
system is rigorously defined; even if its dynamics or other properties are intractable, and 
maybe easier to understand in the limit $N \rightarrow \infty$.

In the foundations and philosophy of quantum field theory, this is a familiar predicament: 
the tension between the  often conflicting virtues of rigour, as in constructive or 
algebraic quantum field theory (AQFT), and heuristic power, as in functional 
(path-integral) quantum field theory. Physicists recognize this tension; although broadly 
speaking, the quantum field theory community is large and liberal enough to accommodate the 
two styles or {\em genres} of work, labelled `mathematical physics' and `theoretical 
physics'.\footnote{\label{Gross}{Agreed, the two styles seem to enjoy a healthy sibling 
rivalry, rather than brotherly love! Cf. Jaffe and Quinn (1993), and Atiyah et al. (1994). 
For a glimpse of the tension, focussed on questions central to this paper, viz. whether 
QCD's asymptotic freedom needs to be proved and is sufficient grounds for believing that 
QCD rigorously exists, we recommend the brief exchange between Jaffe and Gross in Cao 
(1999, pp. 164-165). Jaffe, the mathematical physicist, urges that proofs are needed. The 
theoretical physicist Gross agrees, but emphasizes that he already fully believes QCD 
rigorously exists: if it did not, a problem would surely have already been found. Cf. also 
Gross (1997, pp. 57-62, 1999, p. 571), 't Hooft (1984), and the exchange between 
theoretical physicists at the end of the latter. For a substantial discussion of the 
existence of quantum field theories from the viewpoint of constructive field theory, cf. 
Rivasseau (1991, especially Parts I, III.5).}}

Philosophers of quantum field theory also recognize the tension. Some  say philosophers 
should resist engagement with `merely heuristic' formalisms: the time is not yet ripe for 
philosophical analysis.  Others say that philosophers should ``dive in''. A recent example 
of the debate between these two stances, and how they mould one's assessment of both 
theories and research programmes, is given by papers by Fraser (2011) and Wallace (2006, 
2011).\footnote{{\label{Doreen}}{For a recent general discussion of rigour in physics, cf. 
Davey (2003).}} As is already evident, we are for diving in! We see three reasons for doing 
so.

The first reason is just that despite the lack of rigour in the present-day physics, we 
will be able to classify our three properties in terms of Norton's philosophical taxonomy: 
rigour will not be needed, in order to connect to these philosophical concerns.

The second reason takes longer to state. For it introduces the  notions of asymptotic 
freedom and conformal invariance, which will be crucial in the sequel; and it also leads  
to (i) discussing whether the theories we are concerned with exist in the 't Hooft limit, 
and (ii)
two further philosophical comments. So all this is in Section 2.2.1.A: which will also lead 
to our third reason for diving in (given in Section 2.2.1.B).

\paragraph{2.2.1.A: Asymptotic freedom and conformal invariance}:\\
 Some considerations about quantum field theory suggest that the gauge theories we will 
consider {\em do} exist. Indeed, there are two points here:\\
\indent  (A): a negative one, with a longer history, suggesting that some {\em other} 
quantum field theories do not exist; and \\
\indent  (B): a positive one, with a more recent history, that {\em our} theories probably 
do exist. \\

Broadly speaking, the contrast depends on the type of fields in the theory concerned: 
scalars and fermions are in general subject to (A), while non-abelian gauge fields enjoy 
(B); and theories that combine both types of field need to be analysed individually, case 
by case, to see whether the bad behaviour due to scalars and fermions is outweighed by 
gauge fields' good behaviour. We also take it that this contrast is uncontroversial; cf. 
e.g. Weinberg (1995, Section 18.3, pp. 130-139), Gross (1997, pp. 57-62; 1999, pp. 
573-576), and for a historical introduction, Cao (1999, Sections 10.2-10.3, pp. 290-305). 
We will begin with the negative point, (A).\\

(A): {\em In favour of non-existence}: \\
For many years, various results have suggested that some interacting quantum field 
theories, such as quantum electrodynamics (QED), do not exist, owing to the coupling 
constant becoming infinite at finite energy. For example, Landau thought this sort of bad 
behaviour was generic in quantum field theories; (hence the jargon that the finite energy 
at which the coupling blows up is called `a Landau pole'). This suspicion  has been 
confirmed for certain theories by rigorous non-existence results: Aizenman (1981) proved 
that an interacting scalar quantum field theory in more than four spacetime dimensions, 
with a non-zero $\phi^4$ interacting term, has such behaviour. (For details, cf. Callaway 
(1988, Sections 1-3).)

  Note that this behaviour, the coupling constant becoming infinite, is conceptually 
different from the time-development of an individual solution becoming singular: different, 
and arguably more dismaying, even if such singular solutions are generic, as for example, 
they notoriously are in general relativity. For this behaviour besets even the basic 
dynamical equations of the theory, i.e. its description of physics for arbitrarily short 
times. (This contrast between long and short times is of course crucial throughout physics: 
even in the classical mechanics of point particles, solutions can at long times become 
singular (Xia 1992). Cf. also Witten (1999, p. 1119; 2003, p. 27).)

We should here note one widespread response to this bad behaviour. Namely: we should take 
the theory to be ``just'' phenomenological, describing  accurately a specific (low enough) 
energy-scale.  Generalizing this response to other theories, for higher and higher 
energy-scales, amounts to the effective field theory programme: i.e. the suggestion that 
all our theories are ``just'' phenomenological---each  describing accurately (or at least: 
accurately enough) the physics at some finite range of energies, but each going wrong, or 
even ill-defined, in its description of higher 
energies.\footnote{{\label{Stephan}}{Hartmann (2001) is a good philosophical discussion of 
how such theories can have epistemic virtues, even without having a rigorous mathematical 
formulation. We should also admit that despite what we have said, some distinguished 
mathematicians are working to make sense of theories in case (A): e.g. Connes (2003).}}\\

(B): {\em In favour of existence}: \\
But (A)'s negative results concern quantum field theories that are neither {\em  
asymptotically free} nor {\em conformally invariant}. By briefly explaining these crucial 
properties, it will become clear that a theory with either property might well rigorously 
exist; and we will see later that our theories enjoy one or other property. In fact:  QCD 
has the first, and SYM the second. Both properties arise from the Wilsonian perspective on 
renormalization; i.e. the renormalization group flow systematically redefining coupling 
constants as the energy-scale varies. We will first discuss these matters for finite $N$; 
and then turn to $N = \infty$, i.e. whether there rigorously exist theories QCD($\infty$) 
and SYM($\infty$). This will echo Section \ref{landscape}'s discussion of the schema and 
trichotomy.

{\em Asymptotic freedom}  means that as the energy-scale tends to infinity, the coupling 
constant tends to zero: the particles' interactions die away---they do not feel each other. 
In terms of the renormalization group flow: the beta-function, which controls how the 
coupling constant varies with energy-scale, is negative and so drives the coupling 
constant, and thus itself, to vanish in the limit of infinite energy-scale. In the jargon: 
the theory has an ultra-violet Gaussian fixed point; where `ultra-violet' means 
high-energy, and `Gaussian' means  free, i.e. no interactions. Obviously, this is a very 
striking (Nobel-winning!) feature of a theory. It radically simplifies the physics at 
higher and higher energies: in exactly the regime where we usually fear our quantum field 
theory will break down, the theory becomes free. Thus asymptotically free theories escape 
the above bad behaviour, and the Landau-pole arguments that they do not rigorously exist.

{\em Conformal invariance} means for most theories, including ours, that the theory's 
beta-function is zero at all energy-scales.\footnote{\label{scalecfmal}{This last clause 
really means scale-invariance. But for most theories, this implies conformal invariance; 
and throughout this paper, the difference will not matter.}} Thus there is a fixed point in 
the degenerate sense that the coupling constant does not ``run'', i.e. does not vary with 
energy. Thus the bad behaviour associated with a Landau pole is avoided. We should admit 
that (unlike QCD's being asymptotically free), particle physics is, so far as we know, {\em 
not} conformally invariant. But this of course does not prevent conformal invariance from 
being conceptually important: in particular, by being sufficient for the theory's rigorous 
mathematical existence---as we urge here. Indeed, there are rigorous formulations of some 
conformally invariant theories in two dimensions; (Segal (2004); for an introduction, cf. 
Gannon (2008)).\footnote{Conformal field theories are also conceptually important for other 
reasons: they appear in the world-sheet formulation of string theory, in the AdS/CFT 
correspondence, and in the description of second-order phase transitions (e.g. Cardy 
2008).}

Thus asymptotically free or conformally invariant theories prompt a positive or optimistic 
outlook: such theories probably do rigorously exist. This outlook is well-nigh universal in 
theoretical physics: for example,  recall the views of Gross and 't Hooft about asymptotic 
freedom, reported in footnote \ref{Gross}. Besides, as mentioned in Section \ref{prosp}: 
there are some supersymmetric theories whose beta-function has been computed exactly, and 
some of these are indeed asymptotically free or conformally invariant (Shifman 2012, pp. 
531-533).

Besides, there are various general results relating asymptotic freedom to other important 
features of a theory: so these results are conceptually important. For example: the 
Coleman-Gross theorem says---roughly speaking!---that a necessary condition for asymptotic 
freedom is the use of non-abelian gauge fields. We will glimpse this for QCD in eq. 
\ref{3.2} (in Section \ref{free}): for $SU(N)$, the gauge fields contribute a term 
$\frac{-11}{3}N$ to the beta-function, so that, unless outweighed by other terms dependent 
on the fermions, the function is negative. (For more details, cf. Gross (1999, p. 576-577), 
Zinn-Justin (2002, Chapter 32).) This is certainly one of the main reasons why gauge fields 
are important: if nature ``wants to be well-defined'' by being asymptotically free, then 
she must make use of them!\\

So much by way of generalities. We turn to our theories, QCD and SYM. For QCD, asymptotic 
freedom holds at all finite integers $N$. Usually, this is only stated for the physical 
value $N = 3$; just as we ourselves stated above. But it is easy to argue that it holds for 
all $N$. We will see in Section \ref{free} (again, from eq. \ref{3.2}) that as we let $N$ 
go to infinity in QCD, asymptotic freedom still holds for each value of $N$. Similarly for 
the conformal invariance of SYM. We will see in Section \ref{limitqft} (eq. 
\ref{beta1cfmal}) that as we let $N$ go to infinity in SYM, conformal invariance still 
holds for each value of $N$.\\

{\em The $N = \infty$ cases}: \\
Of course, given the distinction between {\em at} the limit, and {\em on the way} to 
it---more specifically, Butterfield's schema and Norton's trichotomy in Section 
\ref{landscape}---we should ask whether our theories rigorously exist {\em at} the `t Hooft 
limit. As we discussed, the schema allows emergent properties at the limit without 
requiring the existence of the theory {\em at} $N = \infty$. On the other hand, Norton's 
trichotomy, by definition, separated the case (1), where the theory exists, from case (2) 
where it does not. So in order to classify our three properties, (1)-(3) of Section 
\ref{prosp}, in Norton's trichotomy, we need to address this question, whether the theories 
exist at the limit.

As you would expect, this is at present less well-understood than the finite-$N$ situation. 
Nevertheless, we think there are two general reasons to believe they do exist: indeed, 
reasons that also apply more generally to other quantum field theories.\\
\indent (i): First: it is expected that in the 't Hooft limit, some of these theories 
become classical, in the sense that their path-integral reduces to the saddle-point 
approximation; (e.g. Polyakov (1987, Chapter 8.1), Gopakumar and Gross (1994, Section 1)). 
And as we already announced: for SYM, this classicality even gives integrability and a 
correspondence to spin chains (cf. Section \ref{chains}). This integrability is important 
evidence  that SYM rigorously exists in the `t Hooft limit.\footnote{Some other theories 
are also known to be integrable in the 't Hooft limit; the Gross-Neveu model, a theory of 
fermions in 1+1 dimensions, becomes equivalent to an integrable classical sigma-model 
(Gross 1999 p. 585-593).}\\
\indent (ii): Second: there is some pure mathematical understanding of the gauge structure 
of these theories at the 't Hooft limit, especially the Lie algebra $\mathfrak{su}(\infty)$ 
of their group $SU(\infty)$. Cf. Hoppe (1989); or for a review, Rankin (1991, Chapter 2).\\

{\em Two philosophical comments}: \\
 This discussion prompts two further philosophical comments. First: It suggests a 
compromise in the debate among philosophers about whether to dive in to assessing heuristic 
quantum field theory. Namely: there are different cases, i.e. theories, and one should not 
have a uniform stance for all of them. For theories that probably do not exist, as in (A), 
we should treat cautiously their apparent ontological claims, or `world-picture'. So their 
assessment is likely to be a matter of epistemology, not ontology; especially about 
assessing the effective field theory programme mentioned at the end of (A); (cf. footnote 
\ref{Stephan}). But for theories that probably do exist (like QCD and SYM!), 
philosophically assessing their ontology or world-picture is of course worthwhile. And this 
is even true for a theory like SYM that probably does not represent nature, if it helps us 
understand those that do---as SYM is believed to, cf. Section \ref{prosp}.\footnote{This 
last comment is not meant to suggest that we believe that on the other hand, QCD truly 
represents nature at arbitrarily high energies. Of course, we expect that it does not, 
owing to quantum gravity effects at Planck-scale energies. More generally, we endorse the 
consensus that ontology, more generally philosophical interpretation, is worthwhile even 
for physical theories known not be true in all details and-or in all regimes: a happy 
consensus, since otherwise philosophers of physics would have no work to do. For 
discussion, cf. e.g. Van Fraassen (1991, Chapter 1) and Belot (1998, Section 5).}

In saying this, we do not mean to favour for philosophical discussion the asymptotically 
free and conformally invariant theories, i.e. to favour case (B) over case (A). One of the 
main virtues of the theories in (A) is that they are tractable at the low energy-scales we 
can observe, i.e. perturbation theory is reliable. And correspondingly, the down-side for 
asymptotic freedom in case (B) is the vice called `infra-red slavery': i.e. at low 
energy-scales, the theory is very difficult to calculate perturbatively, due to the strong 
coupling. So one needs non-perturbative techniques---and the 't Hooft limit is one of 
these.

Second: The contrast between (A) and (B) prompts a warning about efforts to mathematically 
understand the usual Higgs mechanism for spontaneous symmetry breaking within rigorous 
quantum field theory: an effort enjoined by some of the last decade's philosophical 
literature (e.g. Earman (2003, Section 3f., p. 338f.; 2004, Section 6f., p. 183f.), Healey 
(2007, p. 172)). Namely: although the pure gauge electroweak theory with gauge group $SU(2) 
\times U(1)$ is asymptotically free, the theory with the Higgs boson (a scalar!) added in 
is probably {\em not} asymptotically free: and so probably does not rigorously exist; (cf. 
Weinberg 1995, pp. 153-154).\\

\paragraph{2.2.1.B: Some meanings of `existence'}:\\
Our third reason for diving in arises from our second reason; but it is broader. Namely: 
the spectrum of degrees of rigour that physics exhibits is itself grist to the mill of 
philosophy! Thus one can envisage different precise meanings that a physicist intends by 
saying that a theory `exists' or `is well-defined'; and with these meanings in hand, a 
survey classifying which theories are presently known to enjoy which of the meanings would  
be philosophically very illuminating. We will not try to formulate such a spectrum of 
meanings and survey of theories. But we will sketch how the discussion so far gives the 
materials to do so. (In the next Section, it will become clear how this bears on the 
classification of our theories' three properties.)

First, it is clear enough what  we have intended by the phrase `rigorous existence': 
existence according to the standards prevalent in axiomatic/algebraic  or constructive 
quantum field theory. A bit more exactly: this means the theory's having a consistent  
model in the sense of the work of Haag and Wightman, and their collaborators.

The other weaker meanings of `existence' or `well-definedness' are versions of the
idea that the quantum field theory has finite behaviour at all energy-scales, especially at 
arbitrarily high energies. We have seen two important ways to make this precise: asymptotic 
freedom and conformal invariance. We should mention a third: {\em asymptotic safety}. This 
means that as the energy-scale tends to infinity, the coupling constant tends to a non-zero 
value. So this notion combines there being a renormalization group flow, as in asymptotic 
freedom, with the limit, i.e. the fixed point, not being Gaussian/free, as in conformal 
invariance. Cf.
Weinberg (1979, pp. 798-809; 1997, p. 249), who suggests that this notion could apply to 
quantized
gravity.\\

\paragraph{2.2.1.C: Conclusion}:\\
So much by way of stating our reasons for enthusiastically diving in: the waters of 
heuristic quantum field theory are exhilarating, rather than murky! But finally, we should 
confess that it is not only the {\em existence} of our theories, QCD and SYM, that remains 
to be rigorously proved, even for finite  $N$. Also: various results about them, which we 
will take in our stride, e.g. that SYM is conformally invariant, and the recent results on 
integrability at infinite $N$ that Section \ref{inty} emphasizes, are {\em not} proven as 
rigorously as one would hope for---and as in, for example, theorems in AQFT. For example, 
these results are usually obtained by computations in an expansion.

Of course, taking such heuristic results in our stride does not mean we fully believe every 
such result can, or will one day, be rigorously proven. Thus to take the case of 
integrability: we already admitted that Section \ref{nonplanar} will consider integrability 
failing at finite $N$. And more generally, we of course concede that integrability of a 
finite $N$ interacting quantum field theory in 3+1 dimensions is surely rare indeed. 
Rather, our taking such results in our stride means two things.\\
\indent (i): The results are secure enough that, we say, the time is right (or ripe 
enough!) for philosophical assessment.\\
\indent (ii): Returning to the question whether our theories, QCD and SYM, exist: we shall 
henceforth assume that they {\em do}---even in the `t Hooft limit. Of course, we are not 
sure. But they probably do: and it would be cumbersome to continually qualify our 
statements by repeating that they might not.  Onward!

\subsubsection{How the properties will be classified}\label{classify}
So we henceforward assume that  QCD and SYM exist, even in the `t Hooft limit. Now suppose 
that taking the limit preserves some property of the finite $N$ theories, in the sense of 
Norton's case (1), idealization. That is:  the limit of the properties of the finite $N$ 
theories is the property of the limit theory. We shall claim that this happy  ``meshing'' 
case holds good for both QCD and SYM, as regards our first two properties, i.e. the 
behaviour of the beta-function and planarity. And planarity illustrates Butterfield's 
emergence with reduction; (cf. Section 2.1.C).\footnote{One last {\em cri de coeur} from 
our over-cautious {\em alter ego}! Suppose that the theories do not rigorously exist in the 
`t Hooft limit, but are ``merely'' asymptotically free/conformally invariant there. Indeed, 
we can cautiously suppose that even at finite $N$, they are merely asymptotically 
free/conformally invariant.
Nevertheless, we shall see that our first two properties for the finite-$N$ theories have a 
limit as $N$ tends to infinity.  (That is: `have a limit', by the standards of the 
heuristic study of asymptotically free/conformally invariant  theories.) So clearly, this 
is Norton's case (2), good approximation. That is: one can use the limit of the property as 
an inexact, but accurate enough, description of those theories in the sequence that have a 
high enough value of $N$. And again, planarity illustrates Butterfield's emergence with 
reduction; cf. (1) of Section 2.1.C.}

Our third property, integrability, fares a little differently. It also emerges in the $N = 
\infty$ limit; and with a vivid representation using spin chains. But so far as is known, 
this does {\em not} follow from the finite-$N$ theories also being integrable: they may 
well not be. Instead, it follows, broadly speaking, from these finite-$N$ theories becoming 
increasingly planar, as $N$ tends to infinity. So in Norton's trichotomy, integrability 
illustrates case (3), which he labels `poor approximation'. But as stressed in Section 
2.1.C, such an example can instantiate Butterfield's emergence with reduction: just because 
reduction can use properties (especially, quantities) of the finite-$N$ theories that are 
not the finite cousins of the reduced limit property. We shall claim that integrability in 
the 't Hooft limit is indeed emergent but reduced: so that here, Norton's label, `poor 
approximation', is unfortunate, since it connotes failure where in fact there is a vivid 
success.

So we now close this Section's long preview of what is to come, by spelling out the last 
two paragraphs; in three comments, (a)-(c).

\indent (a):  Since QCD rigorously exists in the `t Hooft limit, Section \ref{tHooft} will 
illustrate Norton's case (1), i.e. idealization, with two important properties of these 
theories: the beta-function being negative, thus securing asymptotic freedom (Section 
\ref{free}); and planarity (Section \ref{planar}). As mentioned in Section \ref{prosp}, 
planarity will be associated with novelty: the appearance of string-like structures and a 
classical regime---thus illustrating emergence combined with reduction.

\indent (b): When we turn to the supersymmetric theory SYM (Sections \ref{limitqft} and 
\ref{inty}), the classifications of these two properties will be as in (a); (with a minor 
qualification which is clear from (B) of Section \ref{rigour}.A). That is: the 
classifications are as for QCD and its higher $N$ ``cousins'', despite the theory being 
very different. Again, assuming rigorous existence: we have Norton's case (1), 
idealization, and for planarity, Butterfield's emergence with reduction. The minor 
qualification is just that the property of the beta-function to consider is now, not it 
being negative, but it being zero. For SYM is  conformally invariant, which means that its 
beta-function is zero at all energies. So there is no scope or need for argument that it is 
driven to zero at high energies.

\indent (c): In Section \ref{inty}, we will address, for SYM, the property of  
integrability; and as announced, we will see that it obtains in the limit. Indeed, this 
integrability is an important part of our  evidence that SYM rigorously exists in the `t 
Hooft limit. This will again illustrate Butterfield's emergence with reduction. But the 
classification of integrability in Norton's trichotomy will be: case (3), rather than case 
(1), as just reported in (a) and (b) for the beta-function and planarity.

The reason is that it is much easier to ascertain at finite $N$ these first two properties 
than integrability. The property labelled `the beta-function' is really a matter of that 
function being negative or zero (the signatures of asymptotic freedom or conformal 
invariance). And one can ascertain at finite $N$, at least perturbatively, that the 
function is negative or zero. Similarly for planarity. We will see in Section \ref{planar} 
that it means the numerical dominance of those Feynman diagrams that can be drawn on a 
plane without intersecting lines. Again, one can ascertain at finite $N$ the contribution 
to a given process made by such diagrams: the point being  that for any process,  as $N$ 
grows, these diagrams' contribution swamps all other contributions. But we will see in 
Section \ref{nonplanar} that it {\em is} difficult to ascertain  at finite $N$ 
integrability in Section \ref{inty}'s sense. The reason is that this sense is relative to a 
perturbative expansion in the parameter $\lambda$, the `t Hooft coupling: $\lambda := g^2 
N$. That is: at each order in $\lambda$, the system is integrable: to be precise, the 
scaling dimension is given by the eigenvalues of a spin chain's Hamiltonian. But it is 
difficult to carry this approach over to the case of finite $N$.

Finally, this classification of integrability prompts a broader suggestion. Namely: perhaps 
it will be a theory  (or even a fragment of a theory) at {\em infinite} $N$, but not at 
finite $N$, that is computationally tractable---and that we can  hope to exploit to better 
understand the  finite $N$ case. We will briefly return to this in Section \ref{concl}. 
This is not to suggest that the physics community regards focussing on the infinite $N$ 
theory as the unique, or best, approach to taming interacting quantum field theories: of 
course, there are several other approaches.

\section{The `t Hooft limit in QCD}\label{tHooft}

In this Section, we sketch: (i) why one seeks an approximation scheme in QCD (Section 
\ref{free}); and (ii) how `t Hooft's choice of the number $N \equiv N_c$ of colours as the 
parameter of an expansion leads, in the limit of large $N$, to planar diagrams, i.e. 
diagrams without intersecting lines (Section \ref{planar}).

\subsection{The need for an approximation scheme}\label{free}
QCD is a rich and complicated theory. Several of its essential features are still poorly 
understood, within physics---let alone philosophy!  These include confinement, dynamical 
mass generation and chiral symmetry breaking. Besides, confinement means that in the low 
energy regime, QCD becomes strongly coupled; and accordingly, increasingly difficult to 
calculate. As mentioned at the end of Section \ref{rigour}.A, this infra-red slavery is the 
down-side of the fact which we will later focus on: that QCD is asymptotically free, i.e. 
that the coupling constant decreases as the energy increases. For QCD, the best available 
approach to deal with the calculational difficulties is to use numerical simulation on a 
lattice. (Incidentally, here we connect with Norton's examples of lattices and real space 
renormalization; cf. Section \ref{landscape}.B.)

But even apart from  calculating specific problems, QCD is so complicated a theory that we 
cannot expect to obtain exact solutions (Dolan et al. 2003). Therefore, even apart from 
using lattices, we need to find some sort or sorts of approximation scheme.\footnote{Here 
we mean `approximation scheme' in physics' usual wide sense, not Norton's specific sense of 
Section \ref{landscape}.} Since a good approximation scheme is traditionally considered to 
require---and is probably only possible if there is---an appropriate expansion parameter, 
we face the question: what possible expansion parameter does QCD contain?

The ordinary coupling constant is not really  a free parameter in QCD. (Though we will not 
need the details, the reason is that as a result of the renormalisation group flow, the 
coupling constant is absorbed into defining the scale of masses by dimensional 
transmutation.) Indeed, this is one of the most important facts  we know about QCD: not 
least because it is this fact that makes the theory both difficult to calculate and hard to 
understand. Thus the theory has no obvious free parameter that could be used as an 
expansion parameter: it is apparently a theory without parameters. Thus one must hope to 
find a non-obvious free parameter.

Famously, 't Hooft (1974) pointed out that  QCD has a non-obvious candidate for an 
expansion parameter. He suggested that one should generalize QCD, from three colours and an 
$SU(3)$ gauge group, to $N \equiv N_c$
colours and a  $SU(N)$ gauge group.\footnote{For an introduction, we recommend Witten 
(1979, 1980) and Coleman (1985, Chapter 8). Witten (1980) includes a motivating discussion 
of taking the limit of elementary wave mechanics (atomic physics) in $N$ spatial 
dimensions, rather than the usual $N = 3$; and seeing the theory simplify at $N = \infty$. 
`t Hooft's proposal had precedents in 1960s work in statistical mechanics; cf. Brezin and 
Wadia (1993, Chapter 1).} More precisely, he expanded the partition function and the 
correlation functions of a $SU(N)$ gauge theory in powers of $N$; and argued that the 
theory simplifies when the number  $N$ of colours is large. In graphic terms: complicated 
Feynman diagrams at finite $N$ are replaced by much simpler {\em planar} diagrams, and so 
one has to cope with far fewer diagrams---as described in Section \ref{planar}. This has 
proven very useful in lattice QCD (cf. Teper (2009) for a recent review); and it prompted 
the hope that one could solve the theory exactly at $N = \infty$, and then one could better 
understand QCD itself by doing an expansion in $1/N = 1/3$. Although these hopes have not 
yet come true, the results to be surveyed in the rest of this paper are surely progress.

We turn to some details. 't Hooft considers a $SU(N)$ gauge theory, with $N_F$ flavors of
fermions (for short: quarks),  transforming in  the fundamental representation of $SU(N)$. 
There are  $N \equiv N_c$ colours; the gauge bosons (for short: gluons) transform in the 
adjoint representation of $SU(N)$ (dimension = $N$). So QCD is the special case: $N = 3, 
N_F = 3$. The gluon field
is an $N \times N$ traceless hermitian matrix; it is written in terms of $T^A$, the 
generators of $SU(N)$, as $A_\mu=A_\mu^A T^A$. Here, $\mu$ is a spacetime index, $\mu$ = 
0,1,2,3; and the superscript $A$ is an index in the adjoint representation, $A = 
1,2,....N$. The element of $A_{\mu}$ on the $a$th row and $b$th column ($a, b= 1,..., N$) 
is written $(A_{\mu})^a_b$. The matrices $T^A$ are normalized so that $\Tr T^A T^B = 
\frac{1}{2} \delta^{AB}$, and the covariant derivative is:
\begin{equation}
D_\mu = \partial_\mu + i {g\over\sqrt N} A_\mu.
\end{equation}
The coupling constant has been chosen to be $g/\sqrt N$, rather than $g$,
because this will lead to a theory with a sensible (and non-trivial) large $N$
limit. The field strength is
\[
F_{\mu\nu}=\partial_\mu A_\nu - \partial_\nu A_\mu + i {g \over \sqrt N}\left[
A_\mu , A_\nu \right],
\]
and the Lagrangian is, with the usual Dirac gamma-matrices and fermions $\psi_k, k = 1,..., 
N_F$:
\begin{equation}\label{3.1}
L = - \frac{1}{2}\Tr F_{\mu \nu} F^{\mu \nu} + \sum_{k=1}^{N_F}
\bar \psi_k \left( i\, \gamma^\mu D_\mu - m_k \right) \psi_k.
\end{equation}
The large $N$ limit will be taken with the number of flavors $N_F$ fixed. (It is
also possible to consider other limits, such as $N \rightarrow \infty$ with
$N_F/N$ fixed.)

One way to understand the $g/\sqrt{N}$ scaling of the coupling constant, and some features 
in the large $N$ limit, is to look at the one-loop $\beta$-function. For most field 
theories, computing the $\beta$-function that defines the renormalization group flow is (as 
Section \ref{new2} indicated) very difficult. But remarkably, the one-loop contribution 
(first quantum correction) gives a lot of information about the theory's ultra-violet 
behaviour and so its existence (Gross 1999, p. 571).

For our $SU(N)$ gauge theory, the one-loop $\beta$-function $\beta_1$, written in terms of 
the coupling constant $g$ and energy-scale $\mu$, is:\
\begin{equation}\label{betaCasimir}
\beta_1(g) := \mu {d g \over d \mu} = - b_1 {g^3 \over 16 \pi^2},\qquad 
b_1=\frac{11}{3}\,C_2(G) - \frac{4}{3}N_F\, C(R),
\end{equation}
where $C_2(G)$ is a numerical constant characteristic of the representation of the gauge 
group $G$ in which the gluons transform, viz. a quadratic Casimir invariant; and $C(R)$ is 
another Casimir invariant characteristic of the representation in which the fermions 
transform. For gluons in the adjoint representation, $C_2(G) = N$; and for fermions in the 
fundamental representation, $C(R) = 1/2$. With this understanding, Eq. (\ref{betaCasimir}) 
yields

\begin{equation}\label{3.2}
\beta_1(g) =  -  {g^3 \over 16 \pi^2}\Big(\frac{11}{3}N - 
\frac{2}{3}N_F\Big).
\end{equation}
The minus sign in Eq. (\ref{3.2}) is the signal of that remarkable feature, asymptotic 
freedom (as long as the number of fermion flavors $N_F$ is small enough), i.e. that $g$ 
decreases at higher energies, with $\lim_{\mu \rightarrow \infty} g = 0$. We argued in 
Section \ref{rigour} that this feature, asymptotic freedom, is a strong reason to believe 
that gauge theories exist.

We now turn to analysing the behaviour of gauge theories in the planar limit. Equation 
(\ref{3.2}) does not
have a sensible large $N$ limit since the one-loop $\beta$-function, i.e. $\beta_1$, is of 
order $N$. Replacing $g$ by
$g/\sqrt{N}$ in eq.~(\ref{3.2}) (i.e. defining $g' := \sqrt{N}g$ and then writing $g$ for 
$g'$)  gives
\begin{equation}\label{betaeq}
\beta_1 = \mu {d g \over d \mu} = - \left( \frac{11}{3} - \frac{2}{3}\frac{N_F}{N}
\right) {g^3 \over 16 \pi^2} .
\end{equation}
The $\beta$-function equation now has a well-defined limit as $N\rightarrow
\infty$. The $N_F$ term is suppressed by $1/N$. Besides, the limiting formula, i.e. without 
$(2/3)N_F/N$, is the formula for the $\beta$-function for the pure-glue sector of the 
theory. Thus the large $N$ limit for QCD, 
with the coupling constant scaling like $1/\sqrt{N}$, is equivalent to taking
the limit $N \rightarrow \infty$, and asymptotic freedom is preserved.\\

Returning to our project of classifying properties in terms of our schema and Norton's 
trichotomy (Sections \ref{landscape} and \ref{classify}): let us consider the property of 
having a negative $\beta$-function. For $N$ sufficiently large, the first term in eq. 
\ref{betaeq} is negative (indeed, we only need $N > \frac{2}{11}N_F$). So the situation is 
as we announced in (a) of Section \ref{classify}. That is: (i) there is no `novelty', and 
so no emergence, in the limit; yet (ii) since we are assuming that QCD rigorously exists in 
the `t Hooft limit, the negative $\beta$-function illustrates Norton's case (1), i.e. 
idealization.

\subsection{The emergence of planarity}\label{planar}
To understand the theory in the $N \rightarrow \infty$ limit, we now analyse Feynman 
diagrams: we will see that they become planar. For this analysis, we need a simple way  to 
count the powers of $N$ in a given diagram.  't Hooft noticed that this could be done with 
a new kind  of graph with a line for each index $a$ and $b$ in $A_\mu \equiv 
(A_{\mu})^a_b$. So double lines between vertices (rather than the usual one line) keep 
track of colour. The result is that every Feynman diagram in a perturbative expansion of 
the original theory can be written as a sum of 't  Hooft
double line graphs: each double line graph gives a particular colour index
contraction of the given Feynman diagram.




For our purposes, the main point about this double line notation is that one can think of 
each double line graph as a polyhedral surface obtained by gluing
polygons together at the double lines. Since each line has an arrow on it, and
double lines have oppositely directed arrows, one can only construct orientable
polyhedra.\footnote{This last remark follows from the theory's using $SU(N)$. For a theory 
with $SO(N)$, the fundamental representation is a real representation, and the lines in a 
't  Hooft graph would not have arrows; so that in this case, it is possible to construct 
non-orientable surfaces such as (polyhedral approximations to) Klein bottles.}

To compute the $N$-dependence requires counting powers of $N$ from sums over
closed colour index loops, as well as factors of $1/\sqrt{N}$ from the explicit
$N$ dependence in the coupling constant. It is convenient to use a rescaled
Lagrangian to simplify the derivation of the $N$-counting rules. We define
rescaled gauge fields $g A /\sqrt{N} \rightarrow \res{A}$, so that the
covariant derivative is $D_\mu = \partial_\mu + i \res A_\mu$; and rescaled
fermion fields  $\psi \rightarrow \sqrt N \res \psi$.
Then the resulting rescaled Lagrangian from Eq. \ref{3.1} has an overall factor $N$ in 
front.
From this Lagrangian, one can read off the powers of $N$ in any Feynman diagram.

Every vertex gives a factor of $N$, and every quark or gluon propagator gives a factor of 
$1/N$. In addition, every colour index loop gives a factor of $N$, since it represents
a sum over $N$ colours. But now we note that in the double line notation where Feynman 
diagrams
correspond to polygons glued together to form polyhedra, each colour index loop is the 
perimeter of a polygon, and so
corresponds to a face of the polyhedron. Thus one finds that a
connected vacuum diagram (i.e.\ with no external lines) is of order
\begin{equation}\label{3.10}
N^{V-E+F} =: N^\chi,
\end{equation}
where $V$ is the number of vertices, $E$ is the number of edges, and $F$ is the
number of faces. But $\chi := V-E+F$ is a topological invariant,
the famous Euler character; and for a connected orientable surface
\begin{equation}\label{3.11}
\chi = 2 - 2 h - b,
\end{equation}
where $h$ is the number of handles, and $b$ is the number of boundaries; (where handles and 
boundaries are themselves topological invariants). For a sphere and its homeomorphs: $h=0$, 
$b=0$, and so $\chi=2$; for a torus and its homeomorphs, $h=1$, $b=0$, and so $\chi=0$.

A quark is represented by a single line, and so a closed quark loop is a
boundary. Thus every closed quark loop brings a $1/N$ suppression. Besides, the maximum
power of $N$ is two, from graphs with $h=b=0$.\footnote{This corresponds to a vacuum energy 
of order $N^2$:  to be
expected since there are $\mathcal{O}\left(N^2\right)$ gluon degrees of
freedom.} These are connected graphs with
no closed quark loops, and with the topology of a sphere. In a similar way, it can be shown 
that in the pure glue (no quarks) sector of the theory, the 't Hooft graphs with the 
highest power (viz. 2) of $N$, which will dominate as $N$ grows, all have the topology of 
the sphere ($h = 0$).

Now we can see how planarity emerges in the $N\to\infty$ limit. Consider first the 
pure-glue sector, and imagine removing one polygon from
the ``sphere'', so that one obtains a punctured sphere, with one puncture. This can be 
flattened
into a diagram drawn on a flat sheet of paper, with the puncture as the outer perimeter. 
One can then glue back the removed polygon by thinking of it as all the
paper exterior to the diagram. Thus the order $N^2$
graphs are planar diagrams: they can be drawn on the
surface of a sheet of paper without having a gluon ``jump'' over another. That is, all
points where gluon lines cross have to be interaction vertices.

A similar result, that planarity emerges for large $N$, holds good for diagrams that depend 
on quarks. The leading diagrams  are of order $N$, with $h=0$ and $b=1$.\footnote{One
might expect the quark contribution to the vacuum energy to be of order $N$,
since there are $N$ quarks of each flavour: cf. the previous footnote.} It turns out that 
these diagrams have
the topology of a punctured sphere with one puncture, with one of the diagram's quark loops 
forming the boundary of
the puncture. One can then flatten out such a diagram into a planar diagram, as we did for 
gluons: the  single quark loop forms the outer perimeter of the diagram.

To sum up: in the pure-glue sector, the leading diagram has the topology of a sphere, but 
can be flattened onto the plane. And in the quark sector,  the leading diagram is also 
planar, with a single quark loop forming its outer perimeter.\footnote{For simplicity, we 
have discussed only connected diagrams. One can obtain the $N$-dependence of a disconnected 
diagram by multiplying the $N$-dependences of all the connected pieces.}

\subsubsection{Conceptual remarks}\label{conceptual}
After these technicalities, we end this Section with four conceptual remarks: the first 
three about the physics, the fourth about our philosophical project of classifying 
properties. 

(1): First: Planarity in the limit may at first seem a ``merely technical'' property, 
compared with e.g. asymptotic freedom. But not so, for two reasons.\\
\indent \indent  (i): It means that in the limit the  perturbation expansion represents the 
theory completely, in the sense that the amplitudes for non-perturbative effects  are 
suppressed. The heuristic reason is that such effects have amplitudes  $\sim \exp ( -1 / 
g^2 )$; but in the 't Hooft limit, we fix $\lambda := g^2 N$, and then we let $N$ tend to 
infinity with $\lambda$ fixed---so that $\exp ( -1 / g^2 ) \equiv \exp ( -N / \lambda )$ 
tends to zero. This is not to say that the completeness of the perturbation expansion 
representation makes the theory easy to work with. One still has to sum over all planar 
Feynman diagrams (a problem attacked in the 1980s by authors such as 't Hooft and 
Rivasseau). Here lies the promise of SYM's integrability: for SYM, planarity leads very 
fortunately to tractable computation; cf. Sections \ref{limitqft} and \ref{inty}. \\
\indent \indent  (ii): The fact that the dominant diagrams at large $N$ look like 
two-dimensional surfaces prompts the idea that these surfaces could be analysed as the 
propagation in time of a one-dimensional object, i.e. a string. Thus planarity, in addition 
to simplifying the theory, suggests a connection between quantum fields and strings. This 
leads to our second remark.\\

(2): Second: In the 1970s, similar string-like structures were noticed in other contexts. 
For example, in the context of lattice gauge theories, Wilson noticed that the strong 
coupling expansion involves a sum over two-dimensional surfaces, as a result of a 
propagation of  one-dimensional objects, viz. (colour)-electric flux lines (often called 
`Wilson loops').\footnote{For an ontological viewpoint about these lines, cf. Healey (2007, 
Chapter 7). While we agree that any conceptual account of gauge theories must consider 
these lines and other extended objects, we must leave this task for another day.} This led, 
naturally enough, to the suggestion that the string-like structures appearing in the 
different contexts were in fact the same; (though there was of course room for doubt, as 
stressed by Polyakov (2010)). 

Polyakov also suggested that these string-like structures could involve an extra spatial 
dimension. His idea was that there was a crucial difference between the classical and 
quantum cases. Classically, the strings have only transverse oscillations. But after 
quantisation, they acquire an extra longitudinal mode, i.e. a Liouville mode; and this 
Liouville mode turns out to play the role of an extra spatial  dimension. This idea became 
of more significance in later years, with the idea of an emergent spatial dimension in 
string theory, as in the AdS/CFT correspondence (Maldacena (1998)), mentioned in Section 
\ref{prosp}. We will return to this at the end of Section \ref{inty}.\\

(3): Third: Another remarkable feature of the 't Hooft limit, is that gauge invariant 
observables  $\mathcal{O}$ become  c-numbers. More precisely: the correlation function of 
$\mathcal{O}$ at two spacetime points $x$ and $y$ (discussed in Section \ref{prosp}) 
factorizes in the 't Hooft limit: 
\begin{equation}\label{factorise}
\langle \mathcal{O}(x)\mathcal{O}(y)\rangle := \int {\cal D} A_\mu 
e^{-S[A_\mu]}\mathcal{O}(x)\mathcal{O}(y) =  
\langle\mathcal{O}(x)\rangle\langle\mathcal{O}(y)\rangle + O(1/N^2)\, ;
\end{equation}
(where for simplicity, we have written the path-integral definition only for the 
gauge-field $A$, and without normalization). This prompts an analogy with classical field 
theories, with correlation functions being characterised in the classical limit, $\hbar \to 
0$, by $1/N^2$ in eq. \ref{factorise}  now playing the role of $\hbar$; so that $\hbar 
\rightarrow 0$ corresponds to $N \rightarrow \infty$. (For a conceptual discussion, cf. 
Witten (1980, Section IV), Gopakumar and Gross (1995)). \\


(4): Fourth: Returning to our project of classifying properties (Sections \ref{landscape} 
and \ref{classify}): let us consider the property, `planarity', that the theory's Feynman 
diagrams that can be drawn on the plane give a dominant contribution  to any process. Then 
the situation is similar to the end of Section \ref{free}. For $N$ sufficiently large (how 
large depending on the process in question), the planar diagrams dominate; and in the limit 
$N \rightarrow \infty$ only planar diagrams contribute. The striking appearance of 
string-like structures and a classical regime certainly counts as novel, and so emergent, 
although reduced i.e. deduced by taking the limit. And since we are assuming that in the `t 
Hooft limit, QCD rigorously exists,  we have Norton's case (1), an idealization.

\section{Introducing super Yang-Mills theory}\label{limitqft}

\subsection{Introducing the theory}\label{SYM}
We now `change horses', i.e. consider a different theory than QCD and its higher-$N$ 
cousins. Namely: $\NN=4$ maximally supersymmetric Yang-Mills theory, the study of which was 
initiated by Brink et al. (1977). Here, $\NN$ is the number of copies of the supersymmetry 
algebra: {\em not} the number of colours, $N$. However, the theory's gauge group will be 
the familiar $SU(N)$. This theory is often called, for short, `${\cal N}=4$ SYM'; where 
`SYM' stands for `super Yang-Mills'. But we will not consider other values of $\cal N$; so 
we will just write `SYM'.

 We mentioned in Section \ref{prosp} that this theory's $N = \infty$ limit is simpler to 
study than that of QCD, and also exhibits planarity and integrability (details in Section 
\ref{inty}). But there are also two other good reasons to study it.

First, it has various remarkable properties. Broadly speaking, it has a large amount of 
symmetry: which is the origin of the simplicity, planarity and integrability just 
mentioned. More specifically, it is conformally invariant, implying that it has no inherent 
scale.  Classically, many theories are  conformal, e.g.  theories with only massless 
fields; (of course, Maxwell theory is the paradigm example).  But  SYM
stays conformal even at the quantum level.  In particular, its $\beta$-function is believed 
to be zero to all orders in perturbation theory; (cf. Section \ref{beta}). And although QCD 
is not conformal, its being asymptotically free means that at high energies it is close to 
being conformal.    Thus many essential features of high energy gluon scattering---which is 
very relevant for the LHC---can be analysed by studying gauge boson amplitudes in SYM.

Second, this theory is the gauge theory `side' of the best-understood example of the 
gravity/gauge, or AdS/CFT, correspondence. (The gravity side is a certain string theory on 
a cousin of anti-de Sitter space: hence the label, with `CFT' standing for `conformal field 
theory'---here SYM.) In Section \ref{prosp}, we postponed this topic to another paper. So 
suffice it to say here that since Maldacena (1998) introduced this correspondence, it has 
taken centre-stage in  the study both of string theories and of high energy quantum field 
theories (including QCD). Besides, we will see at the end of Section \ref{inty} that the 
significance of its topic, integrability, lies largely in the light this sheds on AdS/CFT.

In this Section, we will first describe the fields that make up SYM, and sketch how they 
lead to a vanishing $\beta$-function and so conformal invariance (Section \ref{beta}). Then 
we discuss the symmetries of SYM, and use them to define a class of operators, {\em  
primaries} (Section \ref{symmies}). In Section \ref{GIopors}, we introduce the {\em single 
trace operators}: these function as `building-blocks' of the gauge invariant operators,  in 
the large $N$ limit. We end the Section by introducing the relation to spin chains, which 
looks forward to Section \ref{inty}'s discussion of integrability.

\subsection{The vanishing $\beta$-function and conformal invariance}\label{beta}

The fields  contained in  SYM are the gauge bosons $A_\mu$,  six massless real scalar 
fields $\phi^I$, $I=1\dots 6$, four chiral fermions $\psi^a_{\al}$ and four anti-chiral 
fermions $\overline\psi_{\dot\al\,a}$, with $a=1\dots 4$.   The indices   $\al,\dda=1,2$  
are the spinor indices of the two independent $SU(2)$ algebras that make up the 
four-dimensional Lorentz algebra.  All fields transform  in the adjoint representation of 
the gauge group, which is $SU(N)$; (unlike Section \ref{tHooft}'s QCD, where the fermions 
transformed in the fundamental representation). The covariant derivative is defined, for 
$\chi$ one of the fields $\phi^I$, $\psi^a_\alpha$, etc. by:
\begin{equation}
\DD_\mu\chi(x) := \partial_\mu\chi(x)-[\Aa_\mu(x),\chi(x)]\,,
\end{equation}
where $\Aa_\mu$ is obtained from $A_{\mu}$ by absorbing coupling constants. The 
corresponding field strength is then written $\FF_{\mu\nu}$; and as usual, we have 
$[\DD_\mu, \DD_\nu] = -\FF_{\mu\nu}$.

The fields that transform covariantly under the gauge group $SU(N)$ include:
the scalars $\phi^I$, the fermions $\psi^a_\al$, $\overline\psi_{\dda\,a}$ and the field 
strengths $\FF_{\mu\nu}$.  Since these fields all live in the adjoint representation, their 
transformation under a gauge transformation is
\begin{equation}\label{gtbringback}
\chi(x)\to\chi(x)+[\ve(x) ,\chi(x)]
\end{equation}
where $\chi(x)$ is the covariant field, and $\ve(x)$ is the generator of the gauge 
transformation. We have explicitly included the space-time dependence of the fields to 
emphasize that this is a local transformation.  %
By applying $\DD_\mu$ to a covariant field $\chi(x)$, we can make other covariant fields 
$\DD_\mu\chi(x)$, etc.; while the gauge connection $\Aa_\mu(x)$ does not transform 
covariantly.

We can use this listing of the field content to discuss the scaling of the theory: in 
particular, conformal invariance. We will now sketch why the one-loop $\beta$-function is 
zero.  For any $SU(N)$ gauge theory, the one-loop $\beta$-function $\beta_1$ is given by
\begin{equation}\label{beta1cfmal}
\beta_1(g) := \mu\frac{\partial 
g}{\partial\mu}=-\frac{g^3}{16\pi^2}\left(\frac{11}{3}\,N-\frac{1}{6}\sum_{i} 
C_i-\frac{1}{3}\sum_j \widetilde C_j\right)\,,
\end{equation}
where the first sum is over all real scalars with quadratic Casimir $C_i$ and the second 
sum is over all Weyl fermions with quadratic Casimir $\widetilde C_j$.  Since all fields in 
SYM are in the adjoint representation, all the Casimirs are $N$.  So mere arithmetic 
implies that, with six real scalars ($i = 1,...,6$) and eight Weyl fermions ($j = 
1,...,8$), we have: $\beta_1(g)=0$.

Going beyond one-loop, the $\beta$-function for SYM was shown to be zero for some higher 
loops; and accordingly, it is now believed that the $\beta$-function vanishes at all loops 
and maybe
non-perturbatively, and hence (cf. footnote \ref{scalecfmal}) that the theory is 
conformally invariant (Brink et al. 1983).\\

Here again we can summarize in terms of our classificatory project. In Section \ref{free}, 
we classified QCD's asymptotic freedom as a Nortonian case (1), since we assume that QCD 
rigorously exists in the 't Hooft limit. Similarly here, for SYM's property of conformal 
invariance. That is, as we announced in (b) of Section \ref{classify}: the classification 
is the same: assuming that SYM rigorously exists in the `t Hooft limit, SYM's having a 
vanishing $\beta$-function is a Nortonian case (1).

\subsection{Dilatation and primary operators}\label{symmies}
As just emphasised, the quantum SYM theory is conformally invariant, so that the Poincar\'e 
symmetry is extended to conformal symmetry in four dimensions, yielding a symmetry group 
$SO(4,2)\simeq SU(2,2)$. There is also the global symmetry group acting on the four copies 
of the supersymmetry algebra, the so-called R-symmetry: this group is $SU(4)\simeq SO(6)$. 
Putting these two together with the supersymmetry generators, we get as the overall 
symmetry group of SYM, the superconformal group in four dimensions: which is the graded Lie 
group $PSU(2,2|4)$. The $P$ stands for `projective'.

Here we meet a crucial contrast with the situation for QCD. Classical QCD (with massless 
fermions in Eq.(2.1)) {\em is} conformally invariant (with the symmetry group: $SO(4,2)$). 
But we have every reason to believe that when it is quantised, the conformal symmetry is 
broken, with only Poincar\'{e} symmetry remaining.\footnote{The general topic of classical 
symmetries being lost after quantization is called `anomalies'.} However for SYM, the big 
symmetry group $PSU(2,2|4)$, including the classical conformal symmetry is unbroken by  
quantum corrections. This puts significant constraints on the quantised theory, and 
provides us with a powerful tool.

In this section we will briefly review some important aspects of the $PSU(2,2|4)$ group, 
and how its structure, especially the bosonic subgroup $SU(2,2) \times SU(4)$, implies the 
existence of operators of minimal scaling dimension, called {\em primary operators}, which 
will be crucial for our discussion of integrability. For simplicity, we will consider just 
the bosonic part of the symmetries.

The conformal group has fifteen generators. Ten generators belong to the Poincar\'e 
algebra, which has four generators $P_\mu$ of space-time translations,  and six generators 
$M_{\mu\nu}$ of the $SO(3,1)\equiv SU(2)\times SU(2)$ Lorentz transformations.  The other 
generators of the conformal algebra are the four generators $K_\mu$ of special conformal 
transformations,  and one generator $D$ of dilatations. These generators satisfy the 
commutation relations
\begin{eqnarray}\label{confalg}
&&[D,P_\mu]=-iP_\mu\, ,\,\,\,\,[D,M_{\mu\nu}]=0\, ,\,\,\,\,[D,K_\mu]=+iK_\mu\, 
,\,\,\,\,[P_\mu,K_\nu]=2i(M_{\mu\nu}-\eta_{\mu\nu}D)\,.\nonumber\\
&&[M_{\mu\nu},P_\lambda]=-i(\eta_{\mu\la}P_\nu-\eta_{\la\nu}P_{\mu})\, ,\qquad
 [M_{\mu\nu},K_\lambda]=-i(\eta_{\mu\la}K_\nu-\eta_{\la\nu}K_{\mu})\, . \end{eqnarray}

The dilatation operator $D$ turns out to play a crucial role in the {\em quantum} structure 
of SYM. While the generators of the Poincar\'{e} subgroup of $PSU(2,2|4)$ do not get 
quantum corrections, the dilatation operator $D$ does:
\begin{equation}\label{dilatationop}
D = D_0 + \delta D (g)\, ,
\end{equation}
where $D_0$ is the classical operator and $\delta D$ is the anomalous dilatation operator 
which depends on the coupling $g$.

Now let $\OO(x)$ be  a local operator in the field theory with scaling dimension $\De$. 
(Recall Section \ref{prosp}'s introduction to the idea of scaling dimension.)
The physical idea of $\De$ is that it is the analogue of the mass in QCD. Technically:
under the rescaling $x\to \la x$,  the operator $\OO(x)$ scales as  $\OO(x)\to 
\la^{-\De}\OO(\la x)$; and the dilatation operator $D$  is the generator of these scalings, 
by which we mean that $\OO(x)\to\la^{-i\,D}\OO(x)\la^{i\, D}$. The dimension $\Delta$ is 
$\Delta_0+\gamma$; with $\Delta_0$  the classical dimension corresponding to the classical 
operator $D_0$ in Eq.(\ref{dilatationop}) and $\gamma$ the anomalous dimension arising from 
quantum corrections corresponding to $\delta D$ (Di Francesco et al. 1997). Thus to find  
the anomalous dimension $\gamma$ of  $\OO(x)$, one considers its two-point correlator  with 
itself:
\begin{equation}\label{2pt}
\langle \OO(x)\overline\OO(y)\rangle\approx \frac{1}{|x-y|^{2\Delta}}\, .
\end{equation}
In Section \ref{anom},  we will give more details about computing anomalous dimensions; but 
we will now sketch how the action of $D$ leads to the idea of primary operators.

The action of the dilation operator $D$ on $\OO(x)$ is
\begin{eqnarray}
[D,\OO(x)]=i\left(-\De+x\,\pdx\right)\OO(x)\,.
\end{eqnarray}
Now we apply $D$ to $[K_\mu,\OO(0)]$ and find, using the Jacobi identity, that
\begin{eqnarray}
[D,[K_\mu,\OO(0)]]=i[K_\mu,\OO(0)]-i\De[K_\mu,\OO(0)]\,.
\end{eqnarray}
Thus, the special conformal generator $K_\mu$ creates from  $\OO(x)$ a new local operator, 
$[K_{\mu}, \OO(x) ]$,  which has dimension $\Delta - 1$.  Aside from the identity operator, 
the local operators  in a unitary quantum field theory must have positive dimension.  
Therefore, if we keep creating new lower-dimensional operators by commuting with the 
special conformal generators, we must eventually reach a barrier where we can go no 
further.  Hence the last operator in this chain, $\wt\OO(x)$, must satisfy
\begin{eqnarray}\label{primcond}
[K_\mu,\wt\OO(x)]=0 \,.
\end{eqnarray}

Furthermore, it can be shown that for a given initial operator $\OO(x)$, all the $K_\mu$ 
lead to the same barrier, i.e. the same $\wt\OO(x)$ obeying Eq.(\ref{primcond}). The  
operator $\wt\OO(x)$ is called {\it primary}. Besides, a similar analysis of the fermionic 
operators leads to analogous primary operators.

To sum up: starting with the primary operator $\wt\OO(x)$, we can build new operators with 
the same dimension or higher, by repeatedly commuting it with $D$ (Di Francesco et al. 
1997).  The higher-dimensional  operators are called {\it descendants} of $\wt\OO(x)$.

\subsection{Gauge invariant operators and a Fock space}\label{GIopors}

We now apply the previous Subsection's discussion to the  operators one actually
encounters in  SYM.
The punchline at the end will be that the anomalous dimension of certain operators, which 
turn out to dominate in the large $N$ limit (called {\em single trace} operators), will, 
for large $N$, be encoded in the Hamiltonian of a spin chain, in which each site carries a 
representation of $SO(6)$, the $R$-symmetry subgroup of $PSU(2,2|4)$.


We recall that the physical observables of a gauge theory are gauge invariant operators. In 
SYM, the local gauge invariant operators are made up of products of traces of  the fields 
that transform covariantly under the gauge group $SU(N)$. These fields include
the scalars $\phi^I$, the fermions $\psi^a_\al$, $\overline\psi_{\dda\,a}$, the field 
strengths $\FF_{\mu\nu}$ and their covariant derivatives.
It is thus clear that the single trace local operator
\begin{equation}\label{traceoperator}
\OO(x)=\Tr[\chi_1(x)\chi_2(x)...\chi_L(x)]\,,
\end{equation}
where the trace is over the internal degree of freedom indices, and $\chi_i(x)$ is one of 
the above covariant fields (with or without covariant derivatives), is itself gauge 
invariant.
We can also build other local gauge invariant operators by taking products of traces.

In Section \ref{inty}, we will take the 't Hooft limit, where  the number of colours $N$ is 
large. This limit has the remarkable property that the scaling dimension of the product of 
single trace operators is equal to the sum of their scaling dimensions, so that all 
information about the spectrum of local operators is determined by the single trace 
operators. Thus, for computing dimensions in this limit, it will suffice to concentrate on 
single trace operators.

Among the many remarkable properties of conformal field theories, one is a subtle 
correspondence between operators and states. (This was worked out in the 1980s; cf. Belavin 
et al. (1983).) An example of this occurs for our single trace operators in SYM: viz. we 
represent operators as states in a Fock space built using bosonic and fermionic creation 
operators. We can build a field $\chi_\ell$ ($\ell = 1, ..., L$) within a single trace 
operator by applying to a ground state $|0\rangle$ appropriate elements drawn from two sets 
of bosonic creation operators $A\dag$, $B\dag$,  and a set of fermionic creation operators 
$C^a\dag$.  Here, we define the operator
\begin{equation}
\CC := A\dag_\al A^\al-B\dag_\dda B^\dda+C^a\dag C_a-2\,;
\end{equation}
and the states that correspond to the actual fields are those states $|\chi\rangle$ in the 
Fock space for which $\CC|\chi\rangle=0$.  This set of states, the image of $\CC$, is 
itself a Fock space, which we denote by $\VV$.  For example, some states satisfying the 
$\CC|\chi\rangle=0$
condition, and the fields they correspond to, are:
\begin{eqnarray} \label{singleton}
(A\dag)^{k+1}(B\dag)^kC^a\dag|0\rangle\ \ \ \ \textrm{corresponds to}&&\ 
\DD^k\psi^{a}\nonumber\\
(A\dag)^{k}(B\dag)^kC^a\dag C^b\dag|0\rangle\ \ \ \ \textrm{corresponds to}&&\ 
\DD^k\phi^{ab}\, .
\end{eqnarray}
All the generators of the total symmetry group, i.e. the superconformal group $PSU(2,2|4)$, 
commute with $\CC$; so the symmetry group preserves the $\CC = 0$ eigenspace.

In Section \ref{chains}, we will return to the projected Fock space $\VV$ in more detail. 
But we can already state the key idea about how all this relates to spin chains. (We will 
postpone our philosophical classification of planarity and integrability until after 
Section \ref{inty}'s details.) For a single trace operator with $L$ arguments, 
$\OO(x)=\Tr[\chi_1(x)\chi_2(x)...\chi_L(x)]$, we consider a spin chain of length $L$, each 
of whose sites carries a representation of the $R$-symmetry group $SO(6)$. Thus each site 
corresponds to one of $\OO$'s arguments. And this correspondence is very informative. Not 
only do we have: a state in the Fock space at site $\ell$, built up from the vacuum by a 
sequence of creation operators, corresponds to a field $\chi_{\ell}(x)$. But also: the 
anomalous dimensions of single trace local operators will, for large $N$, be encoded in the 
Hamiltonian of the corresponding spin chain.

\section{Integrability at the limit, and the relation to spin chains}\label{inty}

In Section \ref{anom}, we will first outline the computation of anomalous dimensions: 
recall Section 1's old dream of computing a quantum field theory's mass spectrum, and 
Section \ref{symmies}'s introduction to anomalous dimensions.  More precisely: we discuss 
the one-loop anomalous dimensions for a general set of single trace operators; (recall from 
Section \ref{GIopors} that in the large $N$ limit, the single trace operators encode all 
the spectral information). We will see how in the large $N$ limit, the contributions to the 
anomalous dimensions are dominated by diagrams that can be drawn on a plane, like those 
discussed in Section \ref{planar}. 

Then in Section \ref{chains}, we describe the mapping of the system into the problem of 
computing the energies of a certain spin chain with nearest-neighbour interactions. That 
is: the one-loop anomalous dimensions will be given by the eigenvalues of the corresponding 
spin chain's Hamiltonian.

We stress that as usual, to characterise these anomalous dimensions, one adopts a 
perturbative scheme, e.g. an expansion in the 't Hooft coupling.\footnote{Returning to 
Section \ref{rigour}'s discussion whether a quantum field theory exists: note that for 
certain two-dimensional conformal field theories, the anomalous dimension $\gamma$ can be 
computed in terms of a fractal dimension of a random walk by the Schramm-Loewner evolution 
(cf. Cardy (2005)). And the connection between two-dimensional conformal field theories and 
random walks goes far beyond computing anomalous dimensions. It may also provide a way to 
understand rigorously a class of interacting quantum field theories.} As to the history, we 
note that although related ideas about integrability in quantum field theory were discussed 
by Polyakov (1977), the results about SYM in this Section were mainly prompted by the 
seminal work of Minahan and Zarembo (2003).

\subsection{Computing anomalous dimensions}\label{anom}
In this Subsection we will concentrate on the one-loop anomalous dimensions for single 
trace operators composed of scalar fields $\phi^I$ with no covariant derivatives. (In fact, 
all operators can be built from those whose arguments contain no double derivatives.) 
Recall that the anomalous dimension is given by the exponent in the two-point correlator of 
the operator with itself, eq. \ref{2pt}. All scalar fields have classical dimension 1; and 
so for single trace operators made up only of scalar fields with no covariant derivative, 
the classical
dimension of the operator is $L$, the number of arguments i.e. scalar  fields inside the 
trace.

If the coupling constant $g$ is small, then the anomalous dimension $\gamma$ is much 
smaller than the bare dimension $\Delta_0$: $\gamma \ll \Delta_0$. In this case we can 
approximate the correlator in Eq.(\ref{2pt}) as
\begin{equation}\label{2pta}
\langle \OO(x)\overline\OO(y)\rangle\approx 
\frac{1}{|x-y|^{2\Delta_0}}(1-\gamma\,\ln\Lambda^2|x-y|^2)\,,
\end{equation}
where $\Lambda$ is the cutoff scale. The leading, i.e. classical, contribution to this 
correlator, $1/|x-y|^{2\Delta_0}$,  is called the `tree-level contribution'.

Let us now describe what happens as we let $N\to\infty$. We shall see that:\\
 \indent (1): the ideas in Section \ref{planar}, about diagrams that can be drawn on a 
plane coming to dominate the expansion, occur here also  (Section \ref{plandominate}); and 
\\
 \indent (2): the anomalous dimension is given by an operator $\Gamma$---which in Section 
\ref{chains} will be the Hamiltonian of a spin chain (Section \ref{Gamma}).

\subsubsection{Planar diagrams dominate}\label{plandominate}
 We will consider as an example single trace operators for which: (i) all the arguments 
$\chi$ are the same field (so that the number of arguments $L \geq 2$ becomes a power); and 
(ii) the common argument $\chi$ has trace zero. So the operator is  $\Tr[{\chi}^L]$, with 
$L\ge2$ and $\Tr \chi=0$. It is often written $\Psi_L$ for short. We rescale the operator 
as follows:
\begin{equation}\label{PsiL}
\Psi_{L}:= \frac{(4\pi^2)^{L/2}}{\sqrt{L}N^{L/2}}\Tr 
{\chi}^L=\frac{(4\pi^2)^{L/2}}{\sqrt{L}N^{L/2}}\,{{\chi}^{A}}_{B}{{\chi}^{B}}_{C}\dots 
{{\chi}^{\dots}}_{ A}\qquad A,B,C=1, ..., N\,,
\end{equation}
where we have explicitly put in the colour indices.  The prefactors are for normalization 
purposes.  
At tree level, the correlator of a
${\chi}$ field and its conjugate $\overline {\chi}$ is\footnote{We have ignored the fact 
that, because the fields are in the adjoint representation, ${{\chi}^A}_A=0$: this is 
justifiable when we take the large $N$ limit.}
\begin{equation}
\langle {{\chi}^{A}}_{B}(x){{\overline {\chi}}^{\,C }}_D(y)\rangle_{\rm  tree}= 
\frac{{\delta^A}_D{\delta_{B}}^{C}}{4\pi^2|x-y|^2}\,.
\end{equation}

If we now contract $\Psi_L$ with its conjugate $\overline\Psi_L$, then the leading 
contribution to the correlator comes from contracting the individual fields in order, as 
shown in Figure 1 (a) and (b).  The contribution of all such ordered contractions is
\begin{equation}\label{2pt2}
\langle\Psi_{L}(x)\overline \Psi_{L}(y)\rangle_{\rm ordered}= \frac{L 
N^L}{(\sqrt{L}N^{L/2})^2|x-y|^{2L}}=\frac{1}{|x-y|^{2L}}\,.
\end{equation}
The factor of $N^L$ comes from $L$ factors of ${\delta^{A'}}_A{\delta^A}_{A'}=N$, where 
each double set of delta functions is from contractions of neighbouring fields.  The factor 
of $L$ comes from the $L$ ways of contracting the fields in the plane, of which (a) and (b) 
are two examples.
\begin{figure}\label{contr}
\centerline{\includegraphics[width=13cm]{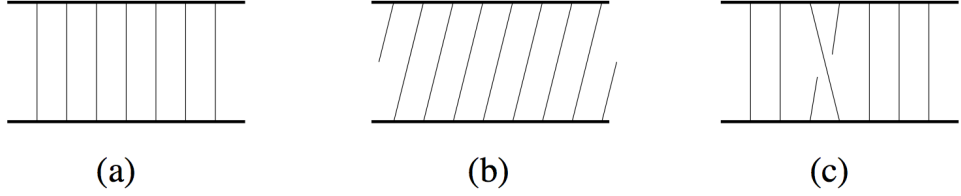}}\caption{The horizontal lines 
represent the operators and the ordered vertical lines represent the contractions between 
the individual fields inside the trace. (a) and (b) are planar; while (c) is nonplanar, 
since lines must cross.} \nonumber
\end{figure}

Figure 1 (c) is an example of a nonplanar diagram, a diagram where the lines connecting the 
fields cannot be drawn in the plane without cutting other lines.  To avoid such cuttings 
one must lift at least one connecting line out of the plane.  The diagram in (c) differs 
from (a) by two field contractions.  Whereas in (a), we would have a factor of
\begin{equation}
\dots{\delta^{A'}}_A{\delta^A}_{A'}{\delta^{B'}}_B{\delta^B}_{B'}{\delta^{C'}}_C{\delta^C}_
{C'}\dots=\dots\,N^3\,\dots\,,
\end{equation}
in (c) we have the factor
\begin{equation}
\dots{\delta^{A'}}_A{\delta^A}_{B'}{\delta^{C'}}_B{\delta^B}_{A'}{\delta^{B'}}_C{\delta^C}_
{C'}\dots=\dots\,N\,\dots\,,
\end{equation}
where the dots represent contractions that are the same in both cases.  Hence, the 
nonplanar diagram in (c) is suppressed by a factor of $1/N^2$ compared to that in (a).   In 
the limit $N\to\infty$, we can thus ignore this contribution compared to the one from (a) 
or (b).

One can show that this example,  $\Psi_L$, is typical. That is: all nonplanar diagrams will 
be suppressed by powers like $1/N^2$, where the power (here, 2) depends on the topology of 
the diagram.\footnote{Actually, this analysis is valid only if $L<<N$.  If $L$ were of the 
order of $N$ then the suppression coming from the $1/N$ factors is swamped by the huge 
number of nonplanar diagrams  compared to the number of planar diagrams.  For there are 
$L!$ total tree-level diagrams of which only $L$ are planar.}\\

Returning to our classificatory project: let us classify the property, planarity, for SYM, 
as we did for QCD in Section \ref{planar}. There are three points to make: the first is an 
echo of our previous classifications, but the second and third will be unlike QCD.\\
\indent (i): First: The situation is as it was for QCD (and as we announced in (b) of 
Section \ref{classify}), in the following sense. We have just seen that as $N$ grows, the 
planar diagrams come to dominate; and in the limit $N \rightarrow \infty$, only planar 
diagrams contribute. So again we have Norton's case (1), assuming that SYM rigorously 
exists in the 't Hooft limit.\\
\indent (ii): Second: Unlike QCD, where we know little about the dynamics of the emergent 
string-like structures, here in SYM we know a good deal about them. For it turns out that 
they are the strings of one of the best-studied string theories, viz. type IIB string 
theory defined on the 10-dimensional spacetime $AdS_5\times S^5$: which is the paradigm 
example of the AdS/CFT correspondence mentioned in Section \ref{prosp}. This returns us to 
the discussion of strings within QCD in remark (2) in Section \ref{conceptual}. The 
dynamics of these required the introduction of extra dimensions (recall Polyakov's 
Liouville dimension). In the context of AdS/CFT, these extra dimensions correspond to the 
energy scale of the renormalisation group flow; (for a recent discussion, cf. Heemskerk and 
Polchinski (2011)). \\ 
\indent (iii): Third: The emergent planarity in SYM has been shown to be associated with 
integrability---as we will see below. This has not been shown for QCD, although QCD may be 
integrable in the limit---and one of course hopes so. For a recent discussion, cf. Belitsky 
et al. (2004).

\subsubsection{The operator $\Gamma$}\label{Gamma}
Without going into details, we report that the one-loop anomalous dimension $\gamma$ is 
encoded in an  operator, $\Gamma$, whose eigenvalues are $\gamma$. (This arises from 
operator mixing.)

The definition of $\Gamma$ is a sum of terms, which involve: (i) an exchange operator 
$P_{\ell,\ell+1}$ which, as its name suggests, exchanges the flavour indices of the $\ell$ 
and the $\ell+1$ arguments inside the trace in Eq.(\ref{PsiL}); and (ii) a trace operator 
$K_{\ell,\ell+1}$ which contracts the flavor indices of fields at neighbouring arguments. 
To be precise: the action of these operators on a sequence of $\delta$-functions, with 
indices $I$ and $J$ labelling the arguments inside (the trace of) our single trace 
operator, is
\begin{equation}
P_{\ell,\ell+1}\,\delta_{I_1}^{J_1}\dots\delta_{I_\ell}^{J_\ell}\delta_{I_{\ell+1}}^
{J_{\ell+1}}\dots\delta_{I_L}^{J_L}=\delta_{I_1}^{J_1}\dots\delta_{I_\ell}^{J_{\ell+1}}
\delta_{I_{
\ell+1}}^{J_{\ell}}\dots\delta_{I_L}^{J_L}\,.
\end{equation}
and
\begin{equation}
K_{\ell,\ell+1}\,\delta_{I_1}^{J_1}\dots\delta_{I_\ell}^{J_\ell}\delta_{I_{\ell+1}}
^{J_{\ell+1}}\dots\delta_{I_L}^{J_L}=\delta_{I_1}^{J_1}\dots\delta_{I_\ell 
I_{\ell+1}}\delta^{{J_\ell}J_{\ell+1}}\dots\delta_{I_L}^{J_L}\,.
\end{equation}

In particular, for operators made up of scalar fields: $\Gamma$ is given by the formula
\begin{equation}\label{Gammaeq}
\Gamma=\frac{\lambda}{16\pi^2}\sum_{\ell=1}^L(1-C-2P_{\ell,\ell+1}+K_{\ell,\ell+1})\,,
\end{equation}
with $\lambda$ is the `t Hooft coupling, and $C$ a constant arising from a certain set of 
diagrams.
And the possible one-loop anomalous dimensions can then be found by diagonalizing $\Gamma$.

\subsection{Spin chains}\label{chains}
Recall that at the end of Section \ref{GIopors}, we built a field $\chi_{\ell}$ ($\ell = 1, 
..., L$) within a single trace operator with $L$ arguments, by using appropriate creation 
operators, giving a Fock space $\VV$, satisfying a constraint $\CC | \chi \rangle = 0$ and 
preserved by the symmetry group, i.e. the superconformal group $PSU(2,2|4)$.

Indeed, the states in this Fock space, such as (\ref{singleton}),  form an irreducible 
representation of  $PSU(2,2|4)$, called the `singleton'  representation.  However, for the 
single trace operators on which we have focussed since Sec \ref{plandominate} (i.e. a 
traceless common argument $\chi$), this cannot be a representation of all gauge invariant 
operators since all of the fields corresponding to these states are traceless.  Hence we 
will need $L\ge2$ fields inside the trace, leading  to tensor products of the singleton 
representation:
\begin{equation}\label{tensorprod}
\VV_1\otimes\VV_2\otimes\dots\otimes\VV_L, \textrm{with}\,\,\,\, \VV_1\cong \VV_j\cong 
\VV\,.
\end{equation}
And the various generators of $PSU(2,2|4)$ on the tensor product have the general form
\begin{equation}\TT=\sum_{\ell=1}^L\oplus \TT_\ell\,,
\end{equation}
where $\TT_\ell$ is the generator at site $\ell$.  We can also define $\CC$ in this way: 
however the projection is still carried out at each site, i.e. $\CC_\ell 
|\chi_\ell\rangle=0$.  A gauge invariant operator is then mapped to a state in the tensor 
product, but because of the cyclicity of the trace, it must be projected onto only those 
states that are invariant under the shift,
\begin{equation}\label{shift1}
\VV_1\otimes\VV_2\otimes\dots\otimes\VV_L\to\VV_L\otimes\VV_1\otimes\dots\otimes\VV_{L-1}\,
.
\end{equation}

Now we are ready for the punchline. In a very impressive collective effort,\footnote{For a 
review we recommend Beisert et al. (2012) and Beisert (2005).} it has been shown that the 
entire class
of scalar single trace operators of length $L$ can be mapped to a Hilbert space of a spin 
chain, i.e. a tensor product of {\em finite}-dimensional Hilbert spaces
\begin{equation}
\HH_1\otimes\HH_2\dots\otimes\HH_\ell\otimes\dots\otimes\HH_L\,.
\end{equation}
Here each $\HH_\ell$ is the Hilbert space for an $SO(6)$ vector representation.
In other words: the  Hilbert space is that of a one-dimensional spin chain with $L$ sites, 
where at each site  there is an $SO(6)$ vector ``spin".  Because of the cyclicity  of the 
trace, we should include the further restriction that the Hilbert space be invariant under 
the shift
\begin{equation}\label{shift}
\HH_1\otimes\HH_2\dots\otimes\HH_\ell\otimes\dots\otimes\HH_L\to\HH_L\otimes\HH_1\dots
\otimes \HH_{\ell-1}\otimes\dots\otimes\HH_{L-1}\,.
\end{equation}

The operator $\Gamma$ in (\ref{Gammaeq}) acts linearly on this space.
Furthermore, it is Hermitian and commutes with the shift in (\ref{shift}).  It also turns 
out that we can treat $\Gamma$ as a Hamiltonian on the spin chain.    The energy 
eigenvalues then correspond to the possible anomalous dimensions for the scalar operators. 
Since the Hamiltonian commutes with the shift, it is also consistent to project onto 
eigenstates that are invariant under the shift.   Because $P_{\ell,\ell+1}$ and 
$K_{\ell,\ell+1}$ in (\ref{shift}) act on neighbouring fields, the spin chain Hamiltonian 
only has nearest-neighbour interactions between the spins.

Although we will not show it here, the Hamiltonian that corresponds to $\Gamma$ for the 
spin-chain is integrable. Integrability is a subtle idea with various aspects: in 
particular, varying across classical and quantum theories. The core classical idea is that 
the number of conserved charges equals the number of physical degrees of freedom. This even 
applies to systems with infinitely many degrees of freedom, e.g. classical sine-Gordon 
theory: it has an infinite number of conserved charges. For SYM, one needs to take account 
of quantum aspects of integrability; (the idea of quantum integrability was discussed by 
Polyakov (1977)). But in short: an integrable infinite quantum system is described by an 
infinite number of commuting scalar charges (Frishman and Sonnenschein (2010, Chap 5)); and 
SYM's integrability turns out to be related to the classical integrability of the 
corresponding string-like structures described by a sigma model. The situation is 
summarized in Table 1, which is based on Frishman and Sonnenschein (2010, p. 332).

\begin{table}[ht]

\caption{Integrability at the limit}
\centering
\begin{tabular}{l*{2}{c}r}

\hline\hline
$\NN =4$ SYM at the limit & Integrable spin chain \\ [0.5ex] 
\hline
Single trace operator & Cyclic spin chain \\
Field operator & Spin at a site \\
Anomalous dilatation operator $\delta D(g)$ & Hamiltonian  \\
Anomalous dimension $\gamma$ & Energy eigenvalue  \\
 [1ex]
\hline
\end{tabular}
\label{table:nonlin}
\end{table}

Finally, we stress that our discussion has been confined to one loop calculations. Going 
beyond  one loop, one finds that the $n$-loop contribution to the anomalous dimension can 
involve up to  $n$ neighboring fields in an effective Hamiltonian.  Therefore, as $N$ 
increases and the `t Hooft coupling $\lambda$ becomes larger, these longer-range 
interactions become more  important; so that at strong coupling the spin-chain is 
effectively long-range.  In this case, the Hamiltonian is not known above the first few 
loop orders.\\

Again, we summarize in terms of our classificatory project. At first, the classification of 
integrability looks similar to the  classification of the  previous property, planarity. 
That is, one expects to say: if SYM rigorously exists in the 't Hooft limit, integrability 
is novel, emergent and a Nortonian case (1). But beware. Agreed: it is novel, emergent and 
yet reduced. But it is reduced to properties different from its ``cousin-property'', 
integrability at finite $N$.  Thus as we warned in Section \ref{classify} (especially (c)): 
so far as is known at present, integrability is an illustration of Norton's case (3). The 
reason is that it is hard to ascertain at finite $N$ integrability in this Section's sense. 
We will take up this topic in Section \ref{nonplanar}.\\

But first, let us briefly discuss integrability's significance. One main significance is 
that it sheds light on the conjectured AdS/CFT correspondence. Indeed, although our 
exposition has not stressed the fact: most of the results reviewed above have used, or been 
inspired by, string-theoretic ideas and results; and often, ideas and results about 
AdS/CFT. And conversely, there is  hope that these results will help prove the conjectured 
correspondence. It is a strong/weak duality: the strong coupling regime of a conformal 
field theory (CFT) corresponds to a weak, even classical, regime of a gravity theory. So 
one would expect such a correspondence to be very difficult to confirm, since one theory is 
computationally  under control only when the other is not.   But integrability of the 
conformal field theory means we can calculate a good deal about it  at strong 
coupling---and so gather the necessary evidence for the conjecture, at least in the 't 
Hooft limit. Besides, integrability of SYM is especially enlightening, since this CFT 
(together with, on the gravity `side', type IIB string theory on the 10-dimensional 
spacetime $AdS_5\times S^5$) forms the best-understood case of the correspondence.

\indent Though many questions remain open, there is reasonable hope that these 
integrability results will teach us how to go back to the physically relevant case of QCD, 
and finally arrive at the long-sought dual description of it by a string theory. It may 
even take us closer to realizing the quantum field theorist's ultimate goal, unfulfilled 
for more than eighty years: completely understanding an interacting relativistic quantum 
field theory in the four space-time dimensions that we are familiar with.

\section{Aspects of non-integrability before the limit}\label{nonplanar}
As announced already in Section \ref{classify}: the first two of our theories' three 
properties, beta-function behaviour and planarity, do not require a further discussion of 
their behaviour before the limit. But on the other hand, integrability does.

The previous sections have focussed on
one particular case of integrability: the `success story' of solving the
exact planar (i.e. $N = \infty$) sector of  SYM.  While this case
has been at the centre of attention, many investigations have tried to extend integrability 
to: (i) the finite $N$ case;  and (ii) other more realistic theories, like QCD. We will 
briefly report some attempts directed at (i). For brevity, we consider only SYM: we set 
aside (ii), other theories---for which, cf. Belitsky et al. (2004).

One general reason why we would expect integrability to fail at finite $N$ is that 
integrability would seem to imply conservation of particle number in any scattering 
process: which would certainly not be expected in a four-dimensional gauge theory (Frishman 
and Sonnenschein 2010, Chapters 5, 18). But even if, accordingly, one focusses on some 
sector or regime of the finite-$N$ theory, not much is known about integrability. The 
finite-$N$ version of the dilatation generator can  be written down easily enough (at least 
in some sub-sectors and to a certain loop order). But attempts to  diagonalize it have so 
far not revealed any traces of integrability.

We turn to sketching some details about the difficulties involved. As we have seen in 
Section \ref{anom},
planar SYM is described by only one parameter,
the `t Hooft coupling $\lambda$, and planar anomalous dimensions have a
perturbative expansion in terms of this
single parameter. It is this fact that made it possible to search initially for 
integrability in
the planar spectrum, by working order by order in $\lambda$. Thus
the concept of `perturbative integrability' was introduced:
meaning that at $l$ loops, i.e. disregarding terms of order $\lambda^{l+1}$, the planar 
spectrum could be described as an integrable system. 

Accordingly, in studying the question of integrability at finite $N$, i.e. before the `t 
Hooft limit, it is natural to follow a similar perturbative approach, expanding in $1/N$. 
But so far, this approach has not borne fruit. Thus going beyond the `t Hooft limit seems 
to require some non-perturbative way of treating  topologies with more handles etc., e.g. 
tori.

As an example of a simple way of getting evidence whether integrability
persists at finite $N$, one can test for so-called degenerate
parity pairs (Beisert et al. 2003; Section 7). This works as follows.  Parity pairs are 
operators with the same anomalous dimension but opposite
parity, where the parity operation on a single trace operator is defined  by

\begin{equation}
{P}\cdot \mbox{Tr}(X_{i_1}\, X_{i_2}\ldots X_{i_n})=
\mbox{Tr}(X_{i_n}\ldots X_{i_2} \, X_{i_1}).
\end{equation}
(For a multi-trace operator, ${P}$ must act on each of its single trace
components.) In the `t Hooft limit, at one-loop, there are a lot of such
parity pairs. The presence of these degeneracies has its origin
in the integrability of the model. ${\cal N}=4$ SYM theory is parity invariant
and its dilatation generator commutes with the parity operation, i.e.
\begin{equation}
[{D},{P}]=0.
\end{equation}
Note that this only tells us that eigenstates of the dilatation generator
can be organized into eigenstates of the parity operator and nothing about
degeneracies in the spectrum. The degeneracies imply the
existence of an extra conserved charge, ${Q}$, which commutes with the
dilatation generator but anti-commutes with parity, i.e.
\begin{equation}
[{D},{Q}]=0,\hspace{0.7cm} \{{P},{Q}\}=0.
\end{equation}
Acting on a state with ${Q}$, one obtains another state with the opposite parity
but with the same energy. Taking  into account non-planar corrections, we expect that
the degeneracies are lifted. Since parity is still conserved, this  indicates (though of 
course, it does not prove) the disappearance of the
extra conserved charges, and a breakdown of integrability.

 To sum up: it seems that there is little hope for integrability of the spectrum of these 
field theories before the `t Hooft limit, at least  in this simple perturbative sense, i.e. 
by expansion in the coupling constant $\lambda$.

In closing, we mention that accordingly, one turns to consider quantities other than 
anomalous dimensions. These include local operators such as structure constants, and 
non-local ones such as Wilson loops, `t Hooft loops, surface operators and domain walls. 
But the evaluation of structure constants is complicated because of operator mixing. 
Turning to non-local quantities: in fact, before the spin chain integrability that we have 
reviewed was discovered, it was already known that the expectation values of
Maldacena-Wilson loops can in certain cases be expressed in terms of
expectation values of a zero-dimensional integrable matrix model. Besides, this connection 
still constitutes one of the most successful tests, so far, of the AdS/CFT correspondence 
beyond the planar limit. But the relation of Maldacena-Wilson loops to spin chain 
integrability is so far not understood.



\section{Conclusion}\label{concl}
Our project in this paper has been to relate the 't Hooft limit to philosophical discussion 
of inter-theoretic relations. More specifically, we have classified the behaviour in this 
limit of three properties of two theories, QCD and SYM. To classify them, we used a schema 
of Butterfield's and a trichotomy of Norton's. Our verdict was that the properties mostly 
illustrate Butterfield's notion of emergence with reduction; and in  Norton's trichotomy, 
mostly his case (1), called `idealization'. But we will not  here give a longer summary: we 
already gave one, by way of orienting the reader, in Sections \ref{prosp} and \ref{new2}, 
especially Section \ref{classify}.

It is clear that the 't Hooft limit is a rich subject, and we have only scratched the 
surface---or, if you prefer, opened Pandora's box. So we will end by stating two topics we 
have not addressed. The first is physical, the second is philosophical.

First: one might use the 't Hooft limit to understand the mass-gap in quantum gauge 
theories, especially QCD. This is widely recognized to be the most important open 
theoretical problem about these theories (Polyakov (1987, Chapter 1.8), Jaffe and Witten 
(2006), Witten (2003, 2008)). Intuitively speaking, the problem arises from the fact that a 
classical non-abelian gauge theory has solutions describing waves obeying a non-linear 
wave-equation. But, unlike classical electromagnetism with its linear wave-equation, we do 
not see any such waves. The property of the quantum theory that is responsible for this is 
the mass-gap: every excitation of the vacuum has an energy of at least $M$, with $M > 0$. 
So the problem is to understand why the quantum theory has this property, while the 
classical one does not. (This is unlike the origin of mass in the electroweak theory, which 
is usually to add a scalar field, the Higgs boson.) For QCD itself, it has often been 
suggested that the 't Hooft limit may provide our best route for solving the problem 
(Polyakov (1987, Chapter 8.4); Witten (2003, p. 25); cf. 1998).\footnote{Incidentally: here 
lies an analogy with the thermodynamic limit's role in understanding phase transitions. 
There, it is surely a sufficient answer to the sceptic's objection that his boiling kettle 
has finitely many atoms, to say that the limit helps us understand the boiling. Similarly 
here: it is surely a sufficient answer to the sceptic's objection that nature no doubt uses 
only finitely many colours, to say that the 't Hooft limit may help us understand the 
mass-gap.}

Second: there is a cluster of logical and philosophical issues, under  headings such as 
`the equivalence of theories', `analogy' and `duality'. We have not here articulated these 
issues, but it is clear that they have been illustrated in several ways. The obvious main 
case is the correspondence between SYM at $N = \infty$ and spin chains. We have also: (i) 
occasionally compared the 't Hooft limit with the thermodynamic limit, and (ii) 
occasionally mentioned AdS/CFT, which is also known as `gravity-gauge duality' (cf. the end 
of Section \ref{chains}). But a detailed treatment of these issues for the 't Hooft limit 
is work for another day. \\ \\

{\em Acknowledgements}:--- This work was supported by a grant from Templeton World Charity 
Foundation. The opinions expressed in this publication are those of the authors and do not 
necessarily reflect the views of Templeton World Charity Foundation. \\ \\

\newpage

\bibliographystyle{chicago}	
\bibliography{HooftLimit}
\nocite{*}


\end{document}